\documentclass[prc,nofootinbib,amsmath,amssymb,superscriptaddress,groupedaddress,showpacs]{revtex4-1}
\usepackage{epsfig}
\usepackage{dcolumn}
\usepackage{bm}
\usepackage{amssymb}
\usepackage{comment}
\usepackage{amsmath}
\usepackage{array}
\RequirePackage{lineno}
\usepackage [english]{babel}
\usepackage [autostyle, english = american]{csquotes}
\MakeOuterQuote{"}
\usepackage[colorlinks=true]{hyperref}
\usepackage[usenames,dvipsnames,table]{xcolor}
\usepackage{color}

\begin{document}
\title{Measuring jet energy loss fluctuations in the quark-gluon plasma via multiparticle correlations}

\author{Abraham Holtermann}
\author{Jacquelyn Noronha-Hostler}
\author{Anne M. Sickles}
\author{Xiaoning Wang}

\affiliation{\small{\it Department of Physics, University of 
Illinois, Urbana, IL 61801, USA}}

\date{\today}
\begin{abstract}
The quark-gluon plasma (QGP) is a high
temperature state of matter
produced in the collisions of two nuclei at
relativistic energies.
The properties of this matter at short distance scales 
are probed using jets with high transverse momentum ($p_T$) resulting from quarks and gluons scattered
with large momentum transfer in the earliest stages of the collisions.
The Fourier harmonics for anisotropies in the high transverse momentum particle yield, $v_n(p_T)$, indicate the path length dependence of jet energy loss within the QGP. We present a framework to build off of measurements of jet energy loss using $v_n(p_T)$ by characterizing fluctuations in jet energy loss that are currently not constrained experimentally. In this paper, we utilize a set of multivariate moments and cumulants as new experimental observables to measure event-by-event fluctuations in the azimuthal anisotropies of rare probes, and compare them to the azimuthal anisotropies of soft particles. Ultimately, these fluctuations can be used to quantify the magnitude and fluctuations of event-by-event jet energy loss. We relate these quantities to existing multivariate cumulant observables, highlight their unique properties, and validate their sensitivities with a Monte Carlo simulation.

\end{abstract}
\newcommand{\qhat}{\mbox{$\hat{q}$}}
\newcommand{\vn}{\mbox{$v_{n}$}}
\newcommand{\Psin}{\mbox{$\Psi_{n}$}}
\newcommand{\nth}{\mbox{$n$-th}}
\newcommand{\vtwo}{\mbox{$v_{2}$}}
\newcommand{\vthree}{\mbox{$v_{3}$}}
\newcommand{\vfour}{\mbox{$v_{4}$}}
\newcommand{\pt}{\mbox{$p_{\textrm{T}}$}}
\newcommand{\pbpb}{PbPb}
\newcommand{\Qn}{\mbox{$\text{Q}_{n}$}}
\newcommand{\etwo}{\mbox{$\varepsilon_2$}}
\newcommand{\avg}[1]{\left\langle #1 \right\rangle}
\newcommand{\zi}{$\zeta_{1,1;2}(V'_nV_n^*, v_n^2;v_n^2)$ }
\newcommand{\zii}{$\zeta_{2;2}(V'_{n}V_{n}^*;v_{n}^2:)$ }

\maketitle

\section{Introduction}

Sufficient energy densities are reached in collisions at the Relativistic Heavy Ion Collider (RHIC) and the Large Hadron Collider (LHC) to produce the quark-gluon plasma (QGP), a state of hot nuclear matter characterized by the deconfinement of the quarks and gluons that comprise the hadronic matter in atomic nuclei~\cite{Busza:2018rrf}.
This matter is describable as a nearly ideal fluid~\cite{Heinz:2013th,Luzum:2013yya,DerradideSouza:2015kpt} 
and jets traveling through
the medium are suppressed as they interact and exchange energy with the medium  
\cite{Cunqueiro:2021wls}.
Jets arise from the hard scattering
between two partons in the early stages of a collision
and propagate through the QGP. The suppression of jets,
\textit{jet quenching}, has been extensively 
studied~\cite{Cunqueiro:2021wls,Qin:2015srf}
because the high momentum jets are sensitive to short
distance scale processes within the QGP.
Understanding how short-distance scale interactions
give rise to the emergent hydrodynamic properties of
the QGP is a major focus of the experimental programs at RHIC and
the LHC \cite{Citron:2018lsq,Achenbach:2023pba,Arslandok:2023utm}.

Fluctuations in the geometry of the individual nuclei
at the instant of the collision give rise to anisotropies
in the azimuthal angular distribution of hadrons in the 
final state through the hydrodynamic evolution of the
QGP \cite{Takahashi:2009na,Alver:2010gr}. The size of these anisotropies
is characterized by the Fourier coefficients, \vn,
which are defined as:
\begin{equation}\label{eq:1}
\frac{dN}{d\phi} \propto 1+ 2\sum_{n=1}^{\infty} \vn \cos \left[n\left(\phi-\Psin \right)\right]
\end{equation}
where $\phi$ is the azimuthal angle of a particle, and \Psin\ defines the \nth\ order event plane.
These quantities have been measured extensively at RHIC and
the LHC~\cite{STAR:2004jwm,PHENIX:2011yyh,STAR:2013qio,ALICE:2016ccg,CMS:2017xnj,ATLAS:2018ezv}. Measurements of \vn\ have been essential to extractions 
of the QGP shear viscosity to entropy density ratio $\eta/s$~\cite{Bernhard:2019bmu,Nijs:2020roc,JETSCAPE:2020shq}. Additionally,
in order to understand how the collision geometry fluctuates on an
event-by-event basis, the distributions of the \vn\ values have
also been measured at low $p_T$~\cite{ATLAS:2013xzf}, providing insight into fluctuations in the QGP initial state produced in heavy ion collisions. 

The Fourier coefficients $v_n$ are also used to describe the azimuthal anisotropies of more restricted classes of objects, such as jets. Measurements indicate that the magnitude of jet quenching is sensitive to
the path length of the jet through the QGP~\cite{ALICE:2015efi,ATLAS:2021ktw,CMS:2022nsv},
the type of parton from which the jet 
originated~\cite{ATLAS:2022fgb,ATLAS:2023iad},
and the structure of the parton shower~\cite{ALargeIonColliderExperiment:2021mqf,ATLAS:2022vii}.
Measurements of these quantities provide information about the magnitude of the
dependence of energy loss on each quantity, but 
do not provide any information on the role of the fluctuations in each 
energy loss process.
These fluctuations are expected to include
fluctuations related to the initial state geometry (as in 
the soft sector) but could also include fluctuations in the energy
loss process itself. Some discussion of observables
sensitive to these fluctuations has appeared in Ref.~\cite{Betz:2016ayq}.
Similarly, studies of the angular correlations of charmed mesons have
been completed in Ref.~\cite{CMS:2021qqk}. 
Most recently, in Ref.~\cite{Holtermann:2023vwr}, the multiparticle correlation framework was extended
and generalized with the goal of defining experimental observables sensitive to the fluctuations in the \vn\ distributions of hard processes.

Jet quenching fluctuation measurements require very large data samples of events with
particles of interest (POI) (e.g. from jets, heavy-flavor, etc)
and particles from the entire event.
The planned high luminosity data-taking over the coming years at RHIC
and the LHC~\cite{Citron:2018lsq,Achenbach:2023pba} will
allow for the first measurements of these quantities.
The LHC Runs 3 and
4 and ongoing detector upgrades will provide much larger data samples for such measurements than
has been previously available. Finally, the very
large event rate for data from sPHENIX will greatly enhance the ability
to perform these measurements at RHIC~\cite{PHENIX:2015siv}.

In this paper, we show that the framework outlined in Ref. \cite{Holtermann:2023vwr} allows for the construction of observables that can evaluate fluctuations and correlations between azimuthal anisotropies for jets and reference particles. We develop a toy model which can be used to calculate the explicit sensitivities of four distinct observables to fluctuations and correlations in soft and hard particle azimuthal anisotropies. Using these sensitivities, we show how measurements of multiple observables can be used in tandem with our toy model to isolate both fluctuations and correlations between azimuthal anisotropies for particles in the hard sector and soft sector. We anticipate that the capacity of our model and observables to evaluate fluctuations in the azimuthal anisotropies of jets can be instrumental in both theoretical and experimental settings to constrain and validate energy loss properties of the QGP.

Our paper is outlined as follows. In Sec.~\ref{sec:2} we illustrate how multiparticle correlations allow for the approximation of different moments of the event-by-event azimuthal anisotropy distributions for jets and reference particles. We then detail how these moments can be used to construct a variety of observables to measure correlations and fluctuations in jet and reference azimuthal anisotropies. In Sec.~\ref{sec:3} we introduce a bivariate copula toy model to describe the joint distributions of azimuthal anisotropies for jet and reference particles. Then, in Sec. \ref{sec:4} we demonstrate the sensitivity of four different observables to second order fluctuations in jet azimuthal anisotropy and correlations between jet and reference azimuthal anisotropies under this toy model. In Sec.~\ref{sec:5} we discuss the applicability of the methods and observables discussed in this paper to constrain fluctuations and correlations experimentally through the joint measurement of multiple observables. Finally, we discuss our outlook and conclusions in Sec.~\ref{sec:6}.

\section{Multiparticle Correlators}\label{sec:2}

\subsection{Azimuthal Anisotropy}

The fundamental variables in the study of azimuthal anisotropy in heavy ion collisions are $v_n$, the $n$th order Fourier coefficients in the angular distribution of particles within an event:
\begin{equation}\label{eq:v_n}
\frac{d N}{d \phi} \propto 1+2 \sum_{n=1}^{\infty} v_n \cos \left[n\left(\phi-\Psi_n\right)\right]
\end{equation}
where the azimuthal angle $\phi_i$ for each particle $i$ in an event with $N$ particles is measured in relation to the event plane $\Psi_n$; an azimuthal angle at which the distribution of $n\phi_i$ approximately symmetric and reaches a maximum \cite{Bilandzic:2020csw}. When using the symbol $v_n$, we specifically refer to the azimuthal anisotropy coefficients for reference particles, i.e. all charged particles. The angular distribution restricted only to POI can likewise be decomposed into Fourier harmonics $v'_n$:
\begin{equation}\label{eq:v'_n}
\frac{d N'}{d \psi} \propto 1+2 \sum_{n=1}^{\infty} v'_n \cos \left[n\left(\psi-\Psi'_n\right)\right]
\end{equation}
where $N'$ indicates the total multiplicity of POI for a single event, and each POI is at azimuthal angle $\psi_i$. The event plane is labeled $\Psi'_n$, about which the angles $n\psi_i$ are symmetric. 
A useful quantity that encodes information about both the event planes and the azimuthal anisotropy are event \textit{flow vectors:}
\begin{eqnarray}
 V_n &= v_ne^{in\Psi_n}\\
 V'_n &= v'_ne^{in\Psi'_n}
\end{eqnarray}
complex numbers with magnitude $v_n$ and $v'_n$ oriented in the direction of the $n$th order event plane angles $\Psi_n$ and $\Psi'_n$.

Event-by-event fluctuations in $v_n$ and $v'_n$ can be calculated using multiparticle correlators. In heavy ion collisions with large multiplicity, these correlators can be used to evaluate statistical moments of $v_n$, and $v'_n$ ~\cite{Bilandzic:2010jr,Voloshin:2008dg}. 
In Ref.~\cite{Holtermann:2023vwr}, a formalism for measuring multiparticle correlations between POI and reference particles was developed that allows for the construction of event-by-event raw moments with arbitrary dependence on $v'_n$ and $v_n$.\footnote{In Ref. \cite{Holtermann:2023vwr} correlations are described that can take arbitrary dependence on both POI and reference particles from multiple harmonics: generally these correlations could evaluate moments of $v'_n$, $v_n$, $v'_m$, $v_m$, $v'_l$, $v_l,$ etc. These correlations also can be used to construct more complex fluctuation observables as described here in Sec. \ref{observablesection}.}
Using flow vectors, we can write correlations that involve arbitrary numbers of POI and reference particles, and relate them to the values of $V'_n$ and $V_n$ as seen below, 
\begin{equation}\label{eq:multicorr}
 \avg{e^{in(\psi_1 + ... + \psi_{k'} - \psi_{k'+1} - ...\psi_{m'})}e^{in(\phi_1 + ... + \phi_k - \phi_{k+1} - ...\phi_m)}} = ({V'}_n)^{k'}({{V'_n}^*})^{m'}({V_n})^{k}({V_n}^*)^{m}
\end{equation}
where $k'$ and $k$ indicate the POI and reference particles respectively for which the angles are added within the correlation, while $m'$ and $m$ indicate the number of POI and reference particles with angles subtracted in the correlation. Here, the angle brackets $\avg{}$ indicate an average taken over all $k'+m'+k+m$ tuples of unique particle angles within an event in which $k'+m'$ particles are POI and $k + m$ particles are reference particles. 
Note that we must have $k'+k = m' +m$, otherwise the above quantity is isotropic, and will average to zero. 

To study fluctuations in $v_n$, multiparticle particle correlations are often used to evaluate moments of the event-by-event $v_n$ distribution. To evaluate moments of both $v_n$ and $v'_n$, a weighted event-by-event average of Eq.~(\ref{eq:multicorr}) is taken, 
\begin{equation}\label{eq:multicorr2}
 \begin{tabular}{c}
 $\avg{\avg{e^{in(\psi_1 + ... + \psi_{k'} - \psi_{k'+1} - ...\psi_{m'})}e^{in(\phi_1 + ... + \phi_k - \phi_{k+1} - ...\phi_m)}}} = \avg{({V'}_n)^{k'}({{V'_n}^*})^{m'}({V_n})^{k}({V_n}^*)^{m}}$ \\
  $=\frac{\sum_{j}W_j \left[({V'}_n)^{k'}({{V'_n}^*})^{m'}({V_n})^{k}({V_n}^*)^{m}\right]_j}{\sum_{j}W_j }$
 \end{tabular}
\end{equation}
where a weighted average is expressed by double brackets,
and the single set of brackets indicates an expectation value, evaluated on the distribution of events indexed by $j$. Here, $\left[({V'}_n)^{k'}({{V'_n}^*})^{m'}({V_n})^{k}({V_n}^*)^{m}\right]_j$ refers to the quantity $({V'}_n)^{k'}({{V'_n}^*})^{m'}({V_n})^{k}({V_n}^*)^{m}$ evaluated within event $j$.
In this paper, we restrict ourselves to multiparticle correlators that correlate a maximum of six particles total within an event, of which a maximum of two particles are POI. This choice is based on current measurements, which indicate a capacity to measure correlations with two jet-like POI \cite{Bilandzic:2010jr,ALICE:2022smy}. The
11 unnormalized correlators with up to two POI are detailed in Table~\ref{correlators}. Note that for two POI in correlations with more than two total particles, there are two ways to construct the correlation: the angles of the two POI are added, or their difference is taken. Both instances are described in Table \ref{correlators}.
\begin{table}[h!]
\caption{Explicit definition of the multiparticle correlations used in this paper, expressing joint moments of the distribution of $v_n$ and $v'_n$ as correlations between up to six particles, and up to two POI.}
\label{correlators}
\begin{tabular}{c c c c}
\hline\hline
 & \textbf{Two Particles} & \textbf{Four Particles} & \textbf{Six Particles} \\
 \textbf{0 POI} & $\avg{\avg{e^{in(\phi_1 - \phi_2)}}}= \avg{v_n^2}$ & $\avg{\avg{e^{in(\phi_1 + \phi_2 - \phi_3 - \phi_4)}}} = \avg{v_n^4}$ & $\avg{\avg{e^{in(\phi_1 + \phi_2 +\phi_3 - \phi_4 - \phi_5 - \phi_6)}}} = \avg{v_n^6}$ \\
\textbf{1 POI} & $\avg{\avg{ e^{i n\left(\psi_1-\phi_2\right)}}}= \avg{V_n'V_n^*}$ & $\avg{\avg{ e^{i n\left(\psi_1+\phi_2-\phi_3-\phi_4\right)}}}= \avg{V'_nV^*_nv_n^2}$ & $\avg{\avg{e^{in(\psi_1 + \phi_2 +\phi_3 - \phi_4 - \phi_5 - \phi_6)}}} = \avg{V'_nV_n^*v_n^4}$ \\
\textbf{2 POI} & $\avg{\avg{e^{in(\psi_1 - \psi_2)}}}= \avg{{v'_n}^2}$ &
\begin{tabular}{c}
 $\avg{\avg{e^{in(\psi_1 + \psi_2 - \phi_3 - \phi_4 )}}}= \avg{(V'_nV_n^*)^2}$\\
 $\avg{\avg{e^{in(\psi_1 - \phi_2 + \psi_3 - \phi_4)}}}=
 \avg{{v'}_n^2 v_n^2} $ 
\end{tabular}
&
\begin{tabular}{c}
 $\avg{\avg{e^{in(\psi_1 + \psi_2 +\phi_3 - \phi_4 - \phi_5 - \phi_6)}}} = \avg{(V'_nV_n^*)^2v_n^2}$ \\
 $\avg{\avg{e^{in(\psi_1 + \phi_2 -\phi_3 - \psi_4 - \phi_5 - \phi_6)}}} = \avg{{v'}_n^2v_n^4}$
\end{tabular}\\
\hline\hline
\end{tabular}
\end{table}

As shown in Eq.~(\ref{eq:multicorr2}), each correlation can be interpreted as an expectation value for powers of $V_n$ and $V'_n$, across an ensemble of events, meaning that they can be considered to be raw moments of a distribution.
However, since $V_n$, $V'_n$, and their odd powers cannot be directly measured, we opt to use stochastic variables $v_n^2, {v'}_n^2$, and the dot product $V'_nV^*_n$. This choice is consistent with the convention for symmetric cumulants ($SC$), which use stochastic variables $v_n^2$ and $v_m^2$, azimuthal anisotropies at two different harmonics \cite{Bilandzic:2013kga}. 

Raw moments of a distribution often roughly correspond to the product of the averages of their stochastic variables:
\begin{equation}
\avg{X^{\nu_x}Y^{\nu_y}Z^{\nu_z}...} \sim \avg{X}^{\nu_x}\avg{Y}^{\nu_y}\avg{Z}^{\nu_z}...
\end{equation} 
where correlations and decorrelations between the variables above result in a deviation of the above expression from equality. To ensure that each of these moments are comparable, and to highlight correlations between each variable, we can normalize each correlation by the product of the averages of its stochastic variables 
\begin{equation}
 n\avg{X^{\nu_x}Y^{\nu_y}Z^{\nu_z}...} \equiv \frac{\avg{X^{\nu_x}Y^{\nu_y}Z^{\nu_z}...}}{\avg{X}^{\nu_x}\avg{Y}^{\nu_y}\avg{Z}^{\nu_z}...} \sim 1.
\end{equation}
 Here and below, we use the notation $n\avg{}$ to indicate a normalized moment. If $n\avg{X^{\nu_x}Y^{\nu_y}Z^{\nu_z}...} > 1$, we expect significant correlations between the stochastic variables $X,Y,$ and $Z$, while $n\avg{X^{\nu_x}Y^{\nu_y}Z^{\nu_z}...} < 1$ indicates an anticorrelation.

The two-particle correlators are only comprised of one stochastic variable, and normalize trivially to $n\avg{v_n^2} = 1$. Each four-particle correlation correlates two stochastic variables. Likewise, each six-particle correlation correlates three stochastic variables. We demonstrate the normalization scheme for each of the four- and six-particle correlators in Table~\ref{tab:correlators2}. These normalizable multiparticle correlators with dependence up to ${V'_n}^2$ on POI angles can be used to construct a variety of angular correlations \cite{Holtermann:2023vwr}, each with their own sensitivities to moments of the distributions of $v'_n$ and $v_n$. 

\begin{table}
\caption{We show normalization schemes for each four- and six-particle correlation. Note that for two POI, there are two possible normalization schemes, both of which we show here.}
\label{tab:correlators2}
\begin{tabular}{c c c}
\hline\hline
 & \textbf{Four Particles} & \textbf{Six Particles} \\
 \textbf{0 POI} & $n\avg{v_n^4} = \frac{\avg{v_n^4}}{\avg{v_n^2}^2}$ & $n\avg{v_n^6}= \frac{\avg{v_n^6}}{\avg{v_n^2}^3}$ \\
\textbf{1 POI} & $n\avg{V'_nV_n^*v_n^2} = \frac{\avg{V'_nV_n^*v_n^2}}{\avg{V'_nV_n^*}\avg{v_n^2}}$ & $n\avg{V'_nV_n^*v_n^4} = \frac{\avg{V'_nV_n^*v_n^4}}{\avg{v_n^2}^2\avg{V'_nV_n^*}}$ \\
\textbf{2 POI}&
\begin{tabular}{c}
 $n\avg{\left(V'_nV_n^*\right)^2} =\frac{\avg{\left(V'_nV_n^*\right)^2}}{\avg{\left(V'_nV_n^*\right)}^2}$\\ $n\avg{{v'}_n^2v_n^2} =\frac{\avg{{v'}_n^2v_n^2}}{\avg{{v'}_n^2}\avg{v_n^2}}$
\end{tabular}
&
\begin{tabular}{c}
  $n\avg{\left(V'_nV_n^*\right)^2v_n^2} =\frac{\avg{\left(V'_nV_n^*\right)^2}}{\avg{\left(V'_nV_n^*\right)}^2\avg{v_n^2}}$\\ $n\avg{{v'}_n^2v_n^4} =\frac{\avg{{v'}_n^2v_n^2}}{\avg{{v'}_n^2}\avg{v_n^2}^2}$
\end{tabular}\\
\hline\hline
\end{tabular}
\end{table}

\subsection{Observables}\label{observablesection}

 Following Ref.~\cite{Holtermann:2023vwr}, we consider three types of observables to quantify fluctuations and correlations in the stochastic variables $\avg{v_n^2},\avg{V'_nV_n^*}$, and $\avg{{v'}_n^2}$, using the correlations detailed in Table \ref{correlators}. In this paper, we examine the sensitivity of these quantities to fluctuations and correlations in $v'_n$ and $v_n$ to demonstrate the feasibility for these quantities to be used experimentally for the same purpose. Each observable has a "key value" that aids interpretation, which is described in Table \ref{tab:critvals}. Comparing any observable with its key value indicates whether or not the stochastic variables are correlated (the observable exceeds the key value), uncorrelated (the observable is equivalent to the key value), or anticorrelated (the observable's value is lower than the key value). The three types of observables are:

 \begin{itemize}
\item \textbf{Moments}:
The observables we evaluate include each of the 11 raw moments designated in Table \ref{correlators}, which can be combined easily to create normalized moments, detailed in Table \ref{tab:correlators2}. While a raw or normalized raw moment has some standalone interpretive value for quantifying the correlations between variables, they are difficult to interpret and tend to have nonlinear relationships with fluctuation measurements, meaning they are generally not used alone in flow analyses.
However, they are necessary to construct more sophisticated fluctuation observables, such as cumulants, and have nontrivial sensitivities to the fluctuations we seek to evaluate.
\item \textbf{Cumulants}: We evaluate cumulant expansions for each of the eight "higher order" moments corresponding to correlators of either four or six particles in Table~\ref{correlators}, again with up to two particles being POI. Cumulants identify the genuine correlation between a set of stochastic variables $\{x_1,...,x_n\}$ by subtracting away cumulants for every smaller subset of stochastic variables $\{x_{i_1},...,x_{i_m}\} \subset \{x_1,...,x_n\}$. 
The derivation for these cumulant expansions is described in Ref. \cite{Holtermann:2023vwr}, but for up to six-particle correlations, they all take the form of either a symmetric cumulant or their generalization to higher order, an asymmetric cumulant ($ASC$) as defined in Refs. \cite{Holtermann:2023vwr,Bilandzic:2013kga,Bilandzic:2021rgb}: 
\begin{eqnarray}
 &SC(X,Y) = \avg{XY} - \avg{X}\avg{Y}\\
 &ASC(X,Y,Z) = \avg{XYZ} - \avg{X}\avg{YZ} - \avg{Y}\avg{ZW} - \avg{Z}\avg{XY} + 2\avg{X}\avg{Y}\avg{Z}
\end{eqnarray}
where each stochastic variable $X,Y,Z$ can be any stochastic variable $v_n^2, {v'}_n^2,$ or $V'_nV_n^*$. 
The cumulants can be normalized simply as: 
\begin{eqnarray}
 nSC(X,Y) = \frac{SC(X,Y)}{\avg{X}\avg{Y}}\\
 nASC(X,Y,Z) = \frac{ASC(X,Y,Z)}{\avg{X}\avg{Y}\avg{Z}}.
\end{eqnarray}
Finally, we note that since these cumulants involve three or fewer stochastic variables (since we restrict ourselves to correlations of no more than six particles), they are equivalent to the central co-moments of these quantities. The covariance between $X$ and $Y$ is measured by $SC(X,Y)$, and the "co-skewness," the correlation between deviations in the mean for $X,Y,$ and $Z$, is measured by $ASC(X,Y,Z)$. 
\item \textbf{Comparisons of Fluctuations}: We finally define four quantities ($\Gamma$ and $\zeta$ for two and three stochastic variables each) that compare the fluctuations in $v_n^2$ to fluctuations with ${v'}_n^2$ and $V'_nV_n^*$. They were initially described in Ref.~\cite{Holtermann:2023vwr} and are written below as functions of normalized moments: 
\begin{eqnarray}\label{Gamma}
 \Gamma(X,Y;v_n^2,v_n^2) &=& n\avg{XY} - n\avg{v_n^4}\\
 \label{3Gamma}
 \Gamma(X,Y,Z;v_n^2,v_n^2,v_n^2) &=& n\avg{XYZ} - n\avg{v_n^6}\\
 \label{zeta}
 \zeta(X,Y;v_n^2,v_n^2) &=& \frac{n\avg{XY}}{n\avg{v_n^2,v_n^2}}\\
 \label{3zeta}
 \zeta(X,Y,Z;v_n^2,v_n^2,v_n^2) &=& \frac{n\avg{XYZ}}{n\avg{v_n^6}}
\end{eqnarray}
where $X$, $Y$, and $Z$ can be any stochastic variable $v_n^2,{v'}_n^2,$ or $V'_nV^*_n$, so long as the total dependence on ${v'}_n^2 \leq 1$ and the total dependence on $V'_nV_n^* \leq 2$. This constraint allows for 12 unique observables which compare the normalized raw moments for a set of arbitrary stochastic variables to normalized raw moments of the same order in $v_n^2$. As detailed in Table \ref{tab:correlators2}, values larger than 0 displayed by $\Gamma$ indicate more significant fluctuations displayed by the random variables $X,Y,$ and $Z$ than are displayed by $v_n^2$. Negative values indicate that $v_n^2$ displays larger fluctuations than are displayed by $X,Y,$ and $Z$. For $\zeta$ a similar relationship exists: greater fluctuations are displayed by $X,Y,$ and $Z$ than by $v_n^2$ when $\zeta > 1$, and $v_n^2$ displays larger fluctuations than $X,Y$ and $Z$ when $\zeta <1$. 
Additionally, while $\zeta$ is nonlinear and describes differences in the overall magnitude of fluctuations, the difference in central moments between two sets of stochastic variables can be written as a sum of different $\Gamma$ observables. $\Gamma$ and $\zeta$ illustrate the differences and ratios between the four- and six-particle contributions to fluctuations in $v'_n$ and $v_n$. For this paper, we have exclusively focused on correlations within one harmonic, and no more than six particles. 
\end{itemize}

\begin{table}[]
 \centering
 \caption{We display a list of key values and their interpretations for each observable discussed in this section. The stochastic variables $X,Y,$ and $Z$ are arbitrary. }
 \label{tab:critvals}
 \begin{tabular}{p{4cm} p{2cm} p{10cm}}
 \hline\hline
   \textbf{Observable} & \textbf{Key Value} & \textbf{Explanation} \\
   $n\avg{XY}, n\avg{XYZ}$ & 1 & Normalized moments with a value greater than unity indicate that $X,Y,$ and $Z$ are positively correlated, while values of less than unity indicate anticorrelation. A value of exactly one indicates that $X,Y,$ and $Z$ are independent. \\
   nSC$(X,Y),$ nASC$(X,Y,Z)$, & 0 & Both normalized and unnormalized symmetric cumulants or asymmetric cumulants with values greater than 0 indicate genuine positive correlations between $X$ and $Y$, or $X,Y,$ and $Z$ respectively. A value of 0 indicates independence between at least two of $X,Y$ or $Z$, and a negative value for a symmetric cumulant or asymmetric cumulant indicates anticorrelation between $X$ and $Y$, or $X,Y,$ and $Z$.\\
   $\Gamma(X,Y;v_n^2,v_n^2)$ $\Gamma(X,Y,Z;v_n^2,v_n^2,v_n^2)$ & 0 & Positive values of $\Gamma$ indicate that the normalized raw moments $n\avg{XY} > n\avg{v_n^4}$ or $n\avg{XYZ} > n\avg{v_n^6}$, meaning that the fluctuations between $X$ and $Y$, or $X,Y,$ and $Z$ are larger than the second or third order fluctuations between $v_n^2$. $\Gamma = 0$ means that the fluctuations have no difference, and $\Gamma < 0$ indicates that the second or third order fluctuations in $v_n^2$ are greater than the fluctuations displayed by $X$ and $Y$, or $X,Y,$ and $Z$. \\
   $\zeta(X,Y;v_n^2,v_n^2)$, $\zeta(X,Y,Z;v_n^2,v_n^2,v_n^2) $ & 1 & Values of $\zeta$ greater than unity indicate that the overall magnitude of the fluctuations in $X$ and $Y$, or $X,Y,$ and $Z$ are larger than the second or third order fluctuations displayed by $v_n^2$. Likewise, for $\zeta <1$ we have that the fluctuations in $v_n^2$ are larger in magnitude than the fluctuations between $X$ and $Y$, or $X,Y,$ and $Z$\\
 \end{tabular}
 
 \label{tab:my_label}
\end{table}

\section{Modeling Azimuthal Anisotropy}\label{sec:3}

\subsection{Parametrizing Azimuthal Anisotropies}

Using multiparticle correlators, we have identified a way to evaluate event-by-event moments of stochastic variables with varying dependence on $v'_n$ and $v_n$ \cite{Holtermann:2023vwr}, and construct fluctuation observables with them. We will show that these fluctuation observables can evaluate correlations and fluctuations in the \textit{actual values} of $v'_n$ and $v_n$. To do this, we parametrize $v'_n$ and $v_n$ using a bivariate distribution, and by sampling these bivariate distributions, demonstrate how various observables can be used to isolate parameters related to the correlation between $v'_n$ and $v_n$, and the fluctuations within $v'_n$. 

Our approach is not new for the study of fluctuations in the soft sector; in Refs. \cite{Voloshin:2007pc,Voloshin:2008dg}, the authors have used assumptions about initial state geometry to create ansatzes for the $v_n$ distribution, and in Refs. \cite{ATLAS:2013xzf,CMS:2017glf} cumulants and unfolding methods are used to approximate the $v_n$ distribution $p(v_n)$, and then solve for the parameter values used in parametrizations of $p(v_n)$. We extend this approach to the study of fluctuations in $v'_n$ as well as correlations between $v'_n$ and $v_n$, using the cumulants and correlations defined in the previous section.

The parametrizations used in our analysis are validated by their ability to represent simulated values of eccentricity using MCGlauber \cite{Loizides:2017ack}, a heavy ion initial state model. Eccentricity is defined similarly to $v_n$ as the coefficients in the Fourier expansion of the energy density $\rho(r,\varphi)$ of the initial state of a heavy ion collision:
\begin{equation}
 \varepsilon_n e^{i n \psi_n} \equiv-\frac{\int r^n e^{in \varphi} \rho(r, \varphi) r \mathrm{~d} r \mathrm{~d} \varphi}{\int r^n \rho(r, \varphi) r \mathrm{~d} r \mathrm{~d} \varphi}
\end{equation}
where $\varphi$ is the azimuthal angle, and $\psi_n$ is an $n$th order event plane. In Refs.~\cite{Noronha-Hostler:2015dbi,Teaney:2010vd,Gardim:2011xv,Niemi:2012aj,Teaney:2012ke,Qiu:2011iv,Gardim:2014tya}, it is shown that $v_n \propto \varepsilon_n$ is a reasonable approximation for $v_2$ and $v_3$ in central to mid-central collisions. Thus, a distribution that accurately reflects MCGlauber simulations also reflects $p(v_n)$ for $n=2$ or $3$ in central to midcentral collisions.\footnote{As is noted in Refs.~\cite{Renk:2014jja,Denicol:2014ywa}, MCGlauber cannot fully describe the eccentricity and $v_n$ distributions measured from the LHC. However, for the purposes of this analysis, we claim that using MCGlauber as a proxy for $p(v_n)$ is adequate for constraining simple quantities like the mean and standard deviations of $p(v_n)$.}

For our analysis, we consider the Gaussian distribution and the elliptic power distribution. The Gaussian distribution lends itself naturally to this analysis because it easily generalizes into a bivariate distribution with an extra parameter corresponding to the Pearson correlation coefficient $\rho$. Additionally, its fluctuations are typically parameterized by $\sigma$, which, when squared, is the exact value of the variance, $\sigma^2$, of the distribution. Under a Gaussian $\varepsilon_n$ distribution parametrized by the mean $\mu$ and the standard deviation $\sigma$,
\begin{equation}
 p(\varepsilon_n) =\frac{1}{\sigma \sqrt{2 \pi}} e^{-\frac{1}{2}\left(\frac{\varepsilon_n-\mu}{\sigma}\right)^2}
\end{equation}
the following equations represent the second and fourth order moments of $v_n$, using 
\begin{equation}\label{eq:gaussmoments}
 \begin{aligned}
  &\avg{\varepsilon_n^2} = \mu^2 + \sigma^2\\
  &\avg{\varepsilon_n^4} = \mu^4+6 \mu^2 \sigma^2+3 \sigma^4
 \end{aligned}
\end{equation}
 $\avg{(kX)^n} = k^n\avg{X^n}$. In general, a measurement of both of the above moments can be used to solve for $\mu$ and $\sigma$, which will provide information about event-by-event fluctuations in the $v_n$ distribution, since $\sigma^2(v_n)$, can be isolated.

\begin{figure}
\centering
\begin{tabular}{cc}
\includegraphics[width=90mm]{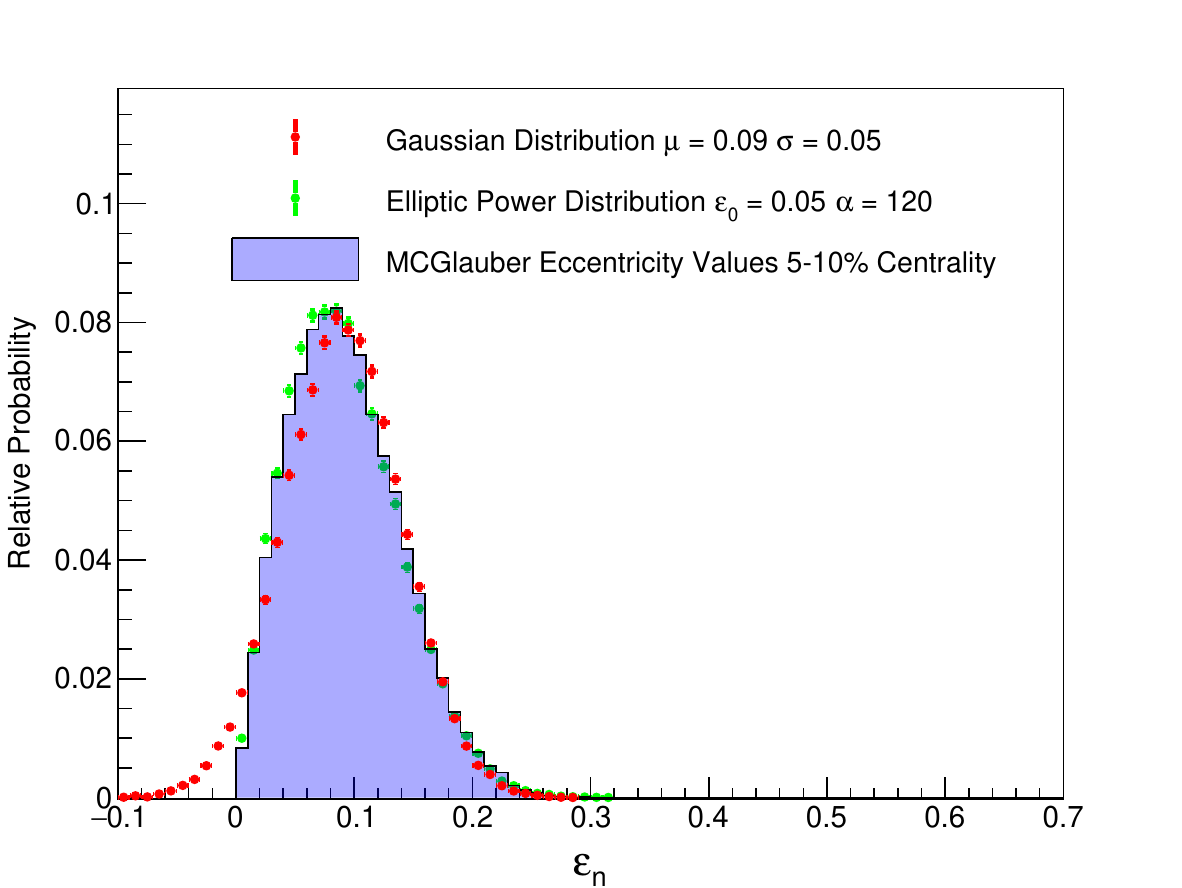} & \includegraphics[width=90mm]{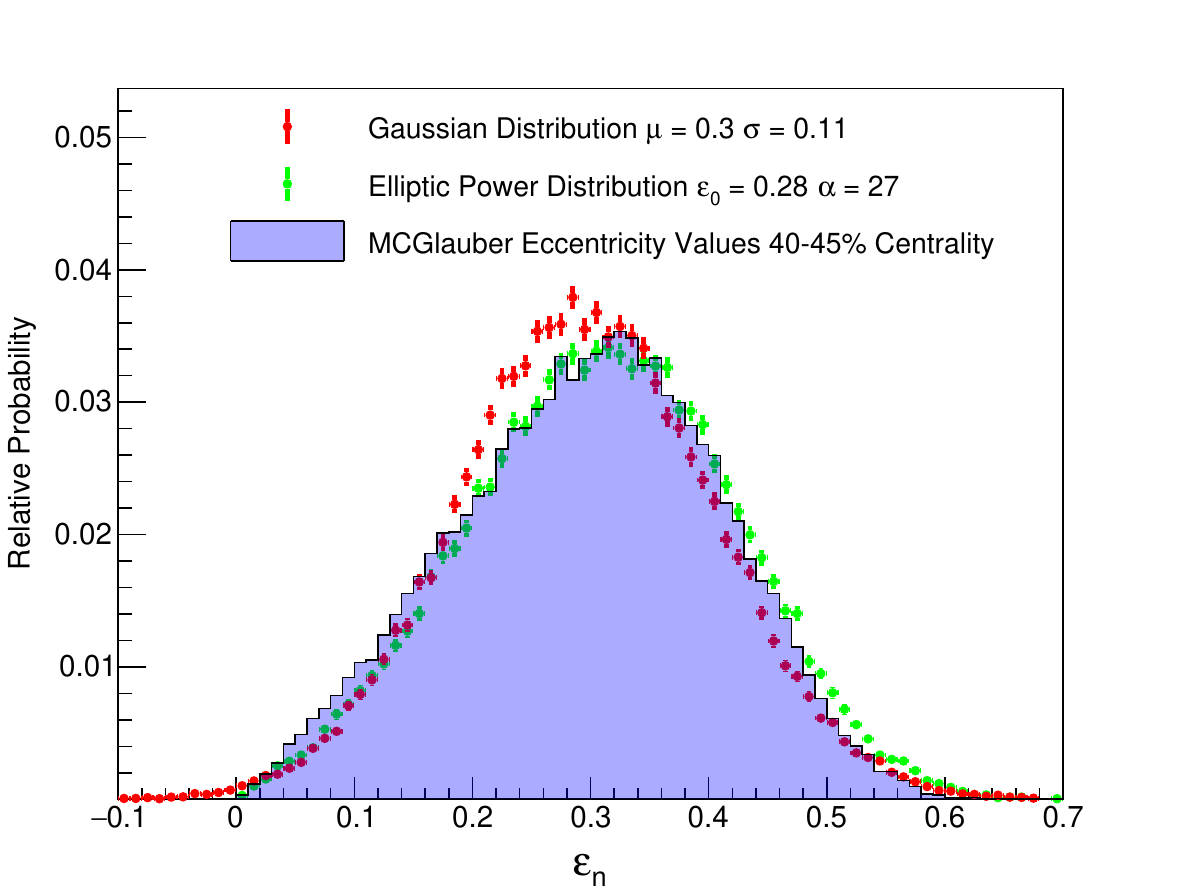}
\end{tabular}
\caption{ Marginal distributions sampled from elliptic power (green) and Gaussian (red) parametrizations are plotted next to MCGlauber eccentricity simulation results at 5-10\% (left) and 40-45\% (right) centrality bins.}
\label{fig:glauberfits}
\end{figure}

The elliptic power distribution was explicitly constructed to represent the eccentricity distributions generated by MCGlauber \cite{Yan:2014afa}. The elliptic power distribution with parameters $\varepsilon_0$, related to the mean $\avg{\varepsilon_n}$, and $\alpha$, related to the spread of $p(\varepsilon_n)$ is written as:
\begin{equation} 
 p(\varepsilon_n)=2 \varepsilon_n \alpha\left(1-\varepsilon_n^2\right)^{\alpha-1}\left(1-\varepsilon_n \varepsilon_0\right)^{-1-2 \alpha}\left(1-\varepsilon_0^2\right)^{\alpha+\frac{1}{2}} { }_2 F_1\left(\frac{1}{2}, 1+2 \alpha ; 1 ; \frac{2 \varepsilon_n \varepsilon_0}{\varepsilon_n \varepsilon_0-1}\right) 
\label{eq:epl}
\end{equation}
where $_2F_1(a,b;c;z)$ is the Gaussian hypergeometric function.
The moments of this distribution are only calculable numerically, but they can be approximated, as shown in Ref. \cite{Yan:2014afa}
\begin{equation}\label{eq:epdcmlt}
 \begin{aligned}
  &\left\langle \varepsilon_n^2\right\rangle\approx\varepsilon_0^2+\frac{\left(1-\varepsilon_0^2\right)\left(1-\frac{3}{2} \varepsilon_0^2\right)}{\alpha}+\mathcal{O}\left(\frac{1}{\alpha^2}\right)\\
  &\left\langle \varepsilon_n^4\right\rangle\approx\varepsilon_0^4+\frac{\varepsilon_0^2\left(1-\varepsilon_0^2\right)\left(4-5 \varepsilon_0^2\right)}{\alpha}+\mathcal{O}\left(\frac{1}{\alpha^2}\right).\\
 \end{aligned}
\end{equation}
 By numerically or analytically estimating the various moments of the elliptic power distribution, we can solve for $\varepsilon_0$ and $\alpha$ in the same way as $\mu$ and $\sigma$ were determined for the Gaussian distribution. Since the relationship between $v_n$ and $\varepsilon_n$ is approximately linear in central to mid-central collisions, a probability distribution $p(\varepsilon_n)$ can also describe the distribution $p(v_n)$. Moreover, due to cancellations, each observable using \textit{normalized} correlators can be calculated using $p(v_n)$ just as well as with $p(\varepsilon_n)$. 
 To indicate the capacity of both the elliptic power and Gaussian parametrizations to characterize realistic event-by-event distributions of $v_n$, sampled values for $\varepsilon_n$ drawn from both elliptic power and Gaussian distributions are shown in Fig.~\ref{fig:glauberfits}. The fit between MCGlauber $\varepsilon_2$ simulated values obtained from Ref.~\cite{Loizides:2017ack}, and $\varepsilon_2$ from the elliptic power and Gaussian distributions is demonstrated for two centrality classes: 5-10\% and 40-45\%. In each plot, the correspondence between the MCGlauber simulation results and the elliptic power and Gaussian distributions is evident, and both parametrizations are shown to accurately characterize significant differences in the fluctuations of $\varepsilon_n$ between the central and peripheral MCGlauber results. 
Note that the Gaussian fit displays nonzero probability for $v'_n$ to have values less than zero. We acknowledge that this is unrealistic. However, it poses no major issues for our analysis, and does not significantly affect our results, as seen by the correspondences in results obtained for both the elliptic power distribution and Gaussian distributions. 

\subsection{Bivariate flow models}

Up until now, we have concentrated on the distribution of $v_n$ values, $p(v_n)$. This distribution is already well studied \cite{ATLAS:2013xzf,CMS:2017glf}, both by approximating the underlying distribution and measuring cumulants. 
In this work, we show that this approach can be extended to the consideration of bivariate flow models that relate $v'_n$ and $v_n$. The distribution $p(v'_n)$ is not currently known, so we introduce a method to construct bivariate azimuthal anisotropy distributions, $p(v_n,v'_n)$ with the same marginal parametrizations for $p(v'_n)$ (although allowing different values and fluctuations) as $p(v_n)$, displaying correlations of varying magnitudes. 

In a previous analysis~\cite{Yi:2011hs}, differential flow $v'_n$ was evaluated using a Gaussian distribution for both $p(v_n)$ and $p(v'_n)$, and assuming a perfect linear correlation between $v_n$ and $v'_n$ to generate $v'_n$ samples by sampling $p(v_n)$~\cite{Yi:2011hs}. While a reasonable approximation, recent measurements from ALICE of \pt\ dependent fluctuations of both the magnitude and the angle
associated with \vtwo\ and $v
'_2$ in \pbpb\ collisions~\cite{ALICE:2022smy} demonstrate that the assumption of perfect correlation is not accurate. In light of this, allowing the correlation coefficient between $v'_n$ and $v_n$ to vary away from unity in our model is necessary to capture realistic decorrelation effects for POI and reference particles. 

\begin{figure}
\includegraphics[width = 18cm]{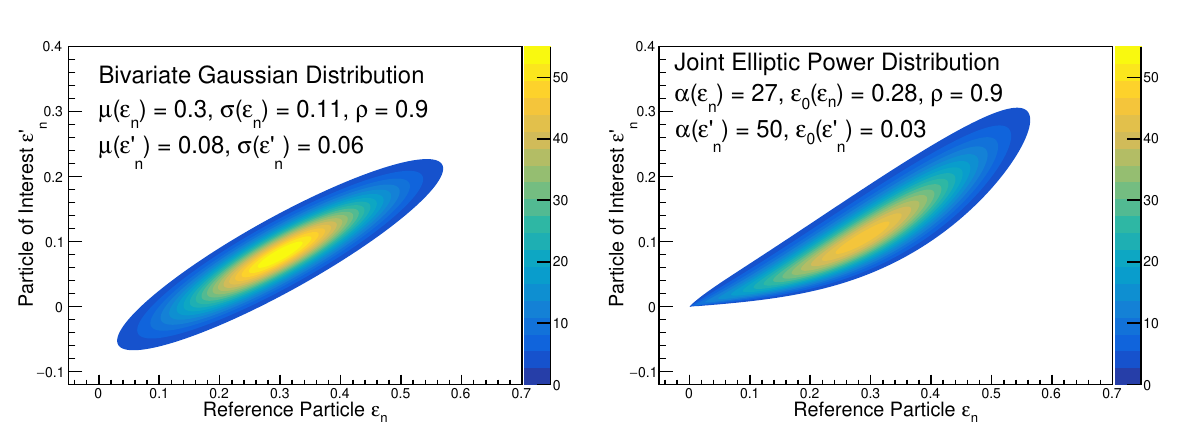}
\caption{Example joint distributions for $\varepsilon_n$ and $\varepsilon'_n$ for the bivariate Gaussian distribution (left) and a joint elliptic power distribution linked with a Gaussian copula (right). The color indicates the magnitude of the probability density for both distributions.}
\label{fig:distributions}
\end{figure}

In this analysis we use the bivariate Gaussian copula \cite{Takeuchi_2010, Meyer_2013} to model the bivariate distribution of $v'_n$ and $v_n$ with a fixed correlation coefficient $\rho$ and marginal distributions $p(v_n)$ and $p({v'}_n)$. Specifically, we examine two cases:
\begin{itemize}
\item \textbf{Bivariate Gaussian Distribution:} The joint probability $p(v_n,v'_n)$ is assumed to have Gaussian marginal distributions, where the standard deviation and mean of the reference $v_n$ distribution are written as $\sigma(v_n)$ and $\mu(v_n)$, and the the same quantities for the $v'_n$ distribution are $\sigma(v'_n)$ and $\mu(v'_n)$.
\item \textbf{Joint Elliptic Power Distribution:} The joint probability $p(v_n,v'_n)$ is assumed to have elliptic power marginal distributions. In this case, the $\varepsilon_0$ and $\alpha$ parameters for reference flow are written as $\varepsilon_0(v_n)$ and $\alpha(v_n)$, and for differential flow, they are written $\varepsilon_0(v'_n)$ and $\alpha(v'_n)$.
\end{itemize}
Using the cumulative distribution function for the Gaussian distribution denoted by $\Phi(x)$: \begin{equation}\Phi(X) = \int_{-\infty}^X \frac{1}{\sqrt{2\pi}}\exp(\frac{-x^2}{2})dx\end{equation}
 we can define the bivariate Gaussian copula density. For a given $\rho$, the bivariate Gaussian copula density $c(v_n,v'_n,\rho)$ is given by \cite{Meyer_2013,Takeuchi_2010}: 
\begin{equation}\label{copula}
 c(v_n,v'_n,\rho) = \frac{1}{\sqrt{1-\rho^2}} \exp \left(\frac{2 \rho \Phi^{-1}(F_1(v_n)) \Phi^{-1}(F_2(v'_n))-\rho^2\left(\Phi^{-1}(F_1(v_n))^2+\Phi^{-1}(F_2(v'_n))^2\right)}{2\left(1-\rho^2\right)}\right)
\end{equation}
where $F_1(v_n)$ and $F_2(v'_n)$ are the respective cumulative distribution functions for the desired marginal distributions of $v_n$ and $v'_n$. The distribution $c(v_n,v'n,\rho)$ is a bivariate probability distribution with marginal distributions characterized by $F_1(v_n)$ and $F_2(v_n)$ and Pearson correlation correlation coefficient $\rho$, that characterizes the probability density for any pair $v_n$, $v'_n$ by their quantile within their marginal distributions, and the selected correlation coefficient. Note that if the marginal distributions are Gaussians, Eq.~(\ref{copula}) reduces to a bivariate Gaussian distribution with correlation parameter $\rho$.
Fig. \ref{fig:distributions} contains example joint distributions for $\varepsilon_n$ and $\varepsilon'_n$ with Gaussian and elliptic power marginal distributions using the same parameter values illustrated in Fig.~\ref{fig:glauberfits} for $p(v_n)$. Both figures have similar mean values and ranges for $v_n$ and $v'_n$. Additionally they illustrate the same values for $\rho$. However, it is clear that constraints imposed by the marginal distributions force qualitative differences between the two distributions, especially for $v_n, v'_n \rightarrow 0$. 

\section{Results}\label{sec:4}

\subsection{Toy Model}

In the previous section we have detailed how to construct bivariate distributions $p(v_n,v'_n)$ using a bivariate copula model. In this section we motivate and describe a toy model using the previously discussed joint distributions to evaluate the sensitivities of the observables developed in Sec. \ref{observablesection} to fluctuations and correlations in $v_n$ and $v'_n$. 

First, given a distribution $p(v_n,v'_n)$, each correlator can be measured by evaluating products of sampled points: $\avg{v_n^2V'_nV_n}$ can be evaluated by simply sampling $p(v_n,v'_n)$ and taking the average value of the product $v_n^3$ and $v'_n$ for each sampled $(v_n,v'_n)$.\footnote{This method relies on the assumption that there is no angular decorrelation between the $n$th order event planes for $\Psi'_n$ and $\Psi_n$. Not only is this a good approximation at low $p_T$ \cite{ALICE:2022smy}, but we can also allow the decorrelations described by $\rho$ in the copula models to describe any desired combination of angular and magnitude decorrelations resulting in a weaker correlation between measured values of the product $v'_nv_n$.} To evaluate the sensitivity of each observable to fluctuations in $v'_n$ and $v_n$, we examine the value of each observable for an array of different parameter values for $p(v_n,v'_n)$. To determine \textit{which} parameters are changed, we consider that both the bivariate Gaussian and joint elliptic power distributions have five parameters: two parameters define $p(v_n)$, two define $p(v'_n)$, and $\rho$ describes the correlation between $v'_n$ and $v_n$. Cumulant analyses as described in Ref.~\cite{Voloshin:2007pc} can constrain both parameters for $p(v_n)$, and differential cumulant studies, as well as scalar product methods \cite{Luzum:2012da,Bilandzic:2010jr} with only one POI can approximate $\avg{v'_n}$, which is typically described by one of the parameters for $p(v'_n)$. This leaves the remaining parameter (the one associated with fluctuations either in $\alpha$ or $\sigma$) for $p(v'_n)$, and $\rho$ to be varied when evaluating the observables in Sec.~\ref{observablesection}. 

\begin{table}
\caption{Parameter values for the bivariate Gaussian and joint elliptic power distributions are displayed. Note that the top three rows are fixed, and the bottom two parameters listed for both distributions are incremented between a range of values.}
\label{tab:params}
\begin{tabular}{c c}
\hline\hline
 \textbf{Bivariate Gaussian } & \textbf{ Joint Elliptic Power} \\
 $\mu(v_n) = 0.3$ & $\varepsilon_{0}(v_n) = 0.28$ \\
$\sigma(v_n) = 0.11$ & $\alpha(v_n) = 27$ \\
$\mu(v'_n) = 0.08$ & $\varepsilon_{0}(v'_n) = 0.03$ \\
$\sigma(v'_n) = 0.8,\cdots, 2.0$ & $\alpha(v'_n) = 3, \cdots, 500$ \\
$\rho = 0.0,\cdots,1.0$&$\rho = 0.0,\cdots,1.0$\\
\hline\hline
\end{tabular}
\end{table}

 Specifically, for the joint elliptic power parametrization, we fix $\varepsilon(v_n),$ $\varepsilon(v'_n)$, and $\alpha(v_n)$, using values obtained for the $40-45\%$ centrality class in Fig.~\ref{fig:glauberfits} before iterating through $\rho$ and $\alpha(v'_n)$. Likewise, for the bivariate Gaussian parametrization, we fix $\mu(v_n)$, $\sigma(v_n)$, and $\mu(v'_n)$, and iterate through values for $\rho$ and $\sigma(v'_n)$. The values for each parameter are found in Table~\ref{tab:params}. 

Our goal is to determine the sensitivity of each observable to fluctuations in $v'_n$ and correlations between $v'_n$ and $v_n$, and show that through the measurement of these observables, correlations between $v'_n$ and $v_n$, and fluctuations in $v'_n$ can be quantified and constrained. To characterize the dependence of each observable on these fluctuations, we require measurements of fluctuations in $v'_n$, and correlations between $v'_n$ and $v_n$. We select $\rho$ as our proxy for correlation between $v'_n$ and $v_n$ because it is among the most commonly used and simplest measurements for correlation between two variables. Additionally, our choice of copula in Eq.~(\ref{copula}) enables for $\rho$ to be selected as a direct input regardless of parametrization.
To evaluate the sensitivity of observables to fluctuations in $v'_n$ we use the relative spread RS: 
\begin{equation}\label{RS}
\mathrm{RS} \equiv\frac{\sigma(v'_n)/\mu(v'_n)}{\sigma(v_n)/\mu(v_n)} 
\end{equation}
which quantifies the fluctuations in $v'_n$ relative to fluctuations within $v_n$ by taking the ratio of their standard deviations divided by their means. This quantity is useful since it is unitless and normalized, and evidently describes fluctuations in $v'_n$ and $v_n$ relative to their mean.

\begin{figure}
 
 \centering
 \includegraphics[width=15cm]{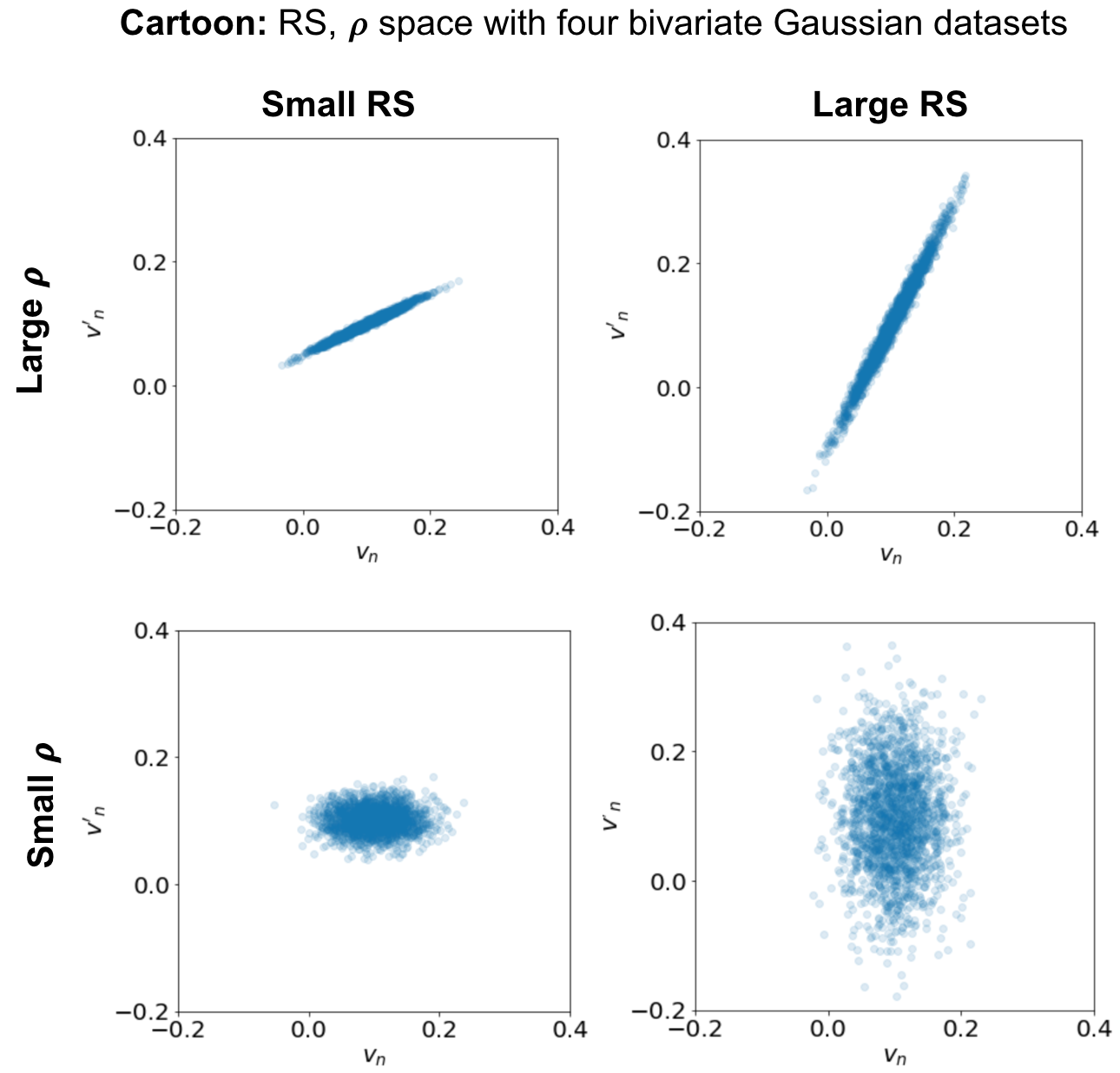}
 \caption{A cartoon illustrating how RS and $\rho$ define a two dimensional space, over which the sensitivity of observables can be evaluated. Example distributions of bivariate Gaussian data distributed according to $\rho$ and RS values are depicted.}
 \label{fig:cartoon}
\end{figure}

RS and $\rho$ are independent quantities, and define a two dimensional space occupied by all bivariate density parametrizations. By evaluating each observable for bivariate distributions at a variety of locations within this space, the sensitivity of each observable to RS and $\rho$ can be identified. Fig.~\ref{fig:cartoon} details this notion, showing example bivariate Gaussian distributions arranged in RS, $\rho$ space. Along the RS axis, the bivariate Gaussian distribution displays increasing values of $\sigma(v'_n)$, and the elliptic power distribution displays increasing values for $\alpha(v'_n)$. Along the $\rho$ axis, $\rho$ changes for both parametrizations. Examples of bivariate Gaussian data for different values of RS and $\rho$ are plotted in the corners, illustrating the qualitative features of the relationship between the bivariate Gaussian distribution, RS and $\rho$, as shown in Fig.~\ref{fig:cartoon}. 

\begin{figure}
 \centering
 \includegraphics[width=10cm]{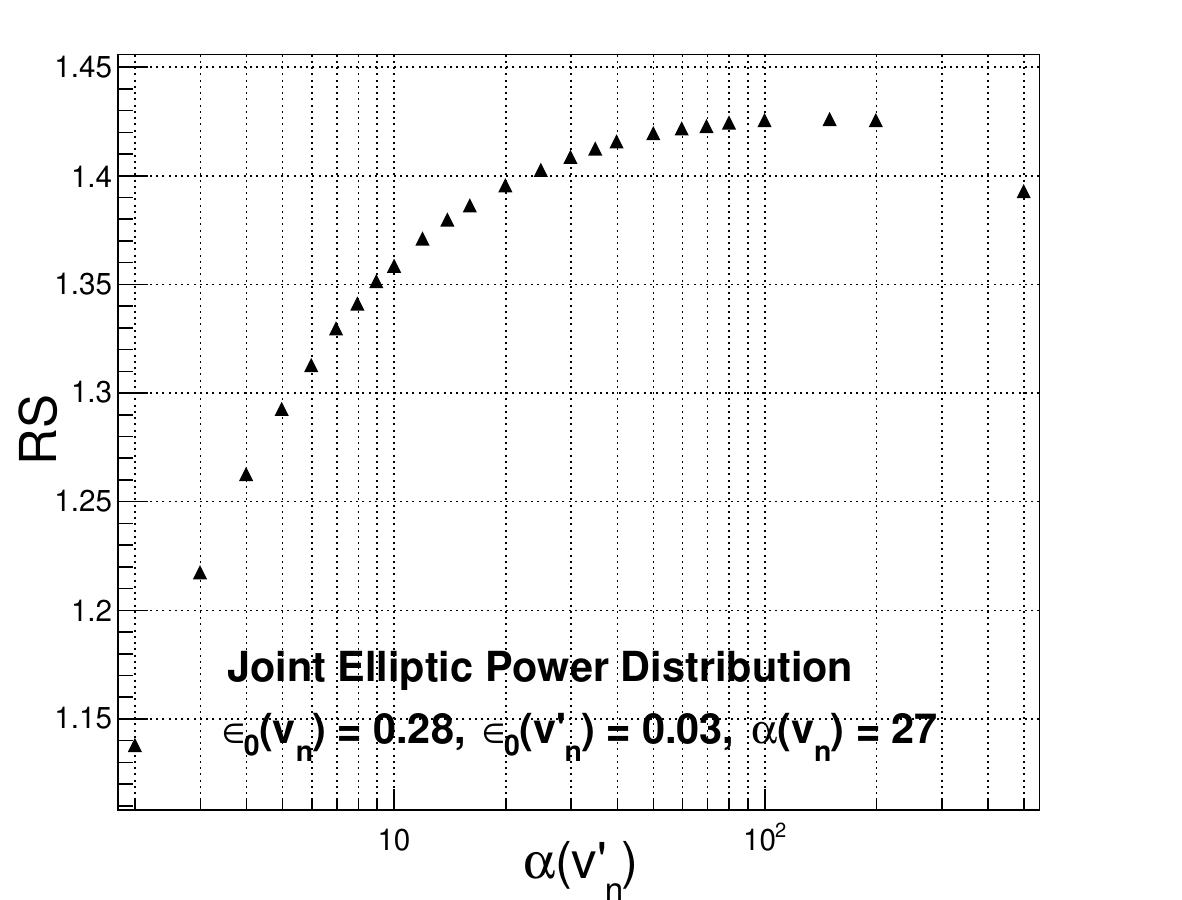}
 \caption{The RS values as defined in Eq.~(\ref{RS}) for the joint elliptic power distribution as a function of $\alpha(v'_n)$ are displayed. The dependence is nonlinear, and is non-monotonic for $\alpha(v'_n) \gg 1$. Our selection of values for $\alpha(v'_n)$ spans the whole range of RS accessible for this parametrization.}
 \label{fig:alphav}
\end{figure}

While RS is easily varied for the bivariate Gaussian distribution, we note that it is more complicated to understand RS in the context of parameters in the elliptic power distribution. Fig. \ref{fig:alphav} shows the correspondence between RS and $\alpha(v'_n)$, where the RS is calculated for the elliptic power distribution at values of $\alpha(v'_n)$ varying from 2 to 500. The fixed values for parameters $\epsilon_0(v_n) = 0.28,$ $ \epsilon_0(v'_n) = 0.03,$ and $\alpha(v_n) = 27$ are selected because they produce the best fit to MCGlauber simulation results for 40-45\% central collisions. From this plot, it is clear that our sampling of $\alpha(v'_n)$ covers the entire range of RS that the joint elliptic power distribution can display with $\epsilon_0(v_n) = 0.28,$ $ \epsilon_0(v'_n) = 0.03,$ and $\alpha(v_n) = 27$.

The code for this toy model is in Ref.~\cite{citation-key}. To run it, a user specifies the desired bivariate flow model, and initial values for the fixed parameters. Then, the code iterates along a defined set of parameter values, at each junction constructing a new bivariate Gaussian or joint elliptic power distribution. The code then samples the newly created distribution ($\sim10^6$ times for the results displayed here), and evaluates $RS$ and $\rho$ by calculating $\rho(v'_n,v_n), \sigma(v'_n),\sigma(v_n), \mu(v_n)$, and $\mu(v'_n)$ from sampled data. The code then evaluates each correlation detailed in Sec.~\ref{observablesection} by averaging products of $v_n$ and $v'_n$. Finally, it evaluates each of the 31 observables using these correlations, and records the value of each observable at its place in the two dimensional space defined by RS and $\rho$.

\subsection{Sensitivity of Four Observables}\label{subsec:4}

Of the observables initially proposed, we present detailed analysis and interpretation of four specific observables that showcase the various types of observable, choices for random variable, and number of particles: $\Gamma(v_n^2,V'_nV_n^{*};v_n^2,v_n^2)$, $\zeta(V'_nV_n^{*},V'_nV_n^{*};v_n^2,v_n^2)$, nSC$({v'}_n^2,v_n^2),$ and $ nASC({v'}_n^2,v_n^2,v_n^2)$. 

We present $\Gamma(v_n^2,V'_nV_n^{*};v_n^2,v_n^2)$ because it requires the least statistical precision, and fewest POIs to obtain a measurement. Additionally, predictions for this quantity are evaluated in Ref. \cite{Betz:2016ayq}, which allows extra interpretive context, indicating different constraints on $\rho$ and RS imposed by various hydrodynamical observables coupled to jet energy loss models. To complement $\Gamma$, which provides a difference between fluctuations with a quantity sensitive to relative fluctuations from four particles, we select $\zeta(V'_nV_n^{*},V'_nV_n^{*};v_n^2,v_n^2)$. This observable has a higher order of dependence on $v'_n$, but correlates the same stochastic variables ($v_n^2$ and $V'_nV_n^*$) as were shown for $\Gamma$. 

As referenced in Sec.~\ref{sec:2}, two-POI correlations have two distinct interpretations: they can correlate $V'_nV_n^*$ and $v_n^2$, or they can correlate ${v'}_n^2$ and $v_n^2$. 
To compare with our study of $V'_nV_n^*$ fluctuations using $\Gamma(v_n^2,V'_nV_n^{*};v_n^2,v_n^2)$ and $\zeta(V'_nV_n^{*},V'_nV_n^{*};v_n^2,v_n^2)$, we use nSC$({v'}_n^2,v_n^2)$ to examine differences in sensitivity to RS and $\rho$ displayed by the stochastic variable ${v'}_n^2$. Finally, to build off our choice for nSC$({v'}_n^2,v_n^2)$, we consider a higher order cumulant using the same variables: nASC$({v'}_n^2,v_n^2,v_n^2)$, to illustrate the viability of higher order cumulants in this analysis, and examine the extent to which higher order cumulants can actually isolate $\rho$ and RS under our toy model.

We evaluate each observable by sampling joint elliptic power distributions and bivariate Gaussian distributions that display varying values of RS, representing the fluctuations in $v'_n$ relative to $v_n$, and $\rho$, representing the correlation between $v'_n$ and $v_n$. In Figs. \ref{fig:deltash} - \ref{fig:nASCs}, we display the sensitivity of $\Gamma,\zeta,nSC,$ and nASC, to RS and $\rho$ using the joint elliptic power and bivariate Gaussian distributions. RS and $\rho$ values are represented for each distribution on the horizontal and vertical axes respectively, while the value of the observable is indicated by color, delineated on a third axis. The level curves (bands within RS, $\rho$ space that are the same color) indicate regions of RS, $\rho$ space over which the observable maintains approximately the same value. In the following text we discuss results for each observable in detail.
 
\subsubsection{ $\Gamma(v_n^2,V'_nV_n^{*};v_n^2,v_n^2)$}

\begin{figure}
 \centering
 \begin{tabular}{ll}
 \includegraphics[width=9cm]{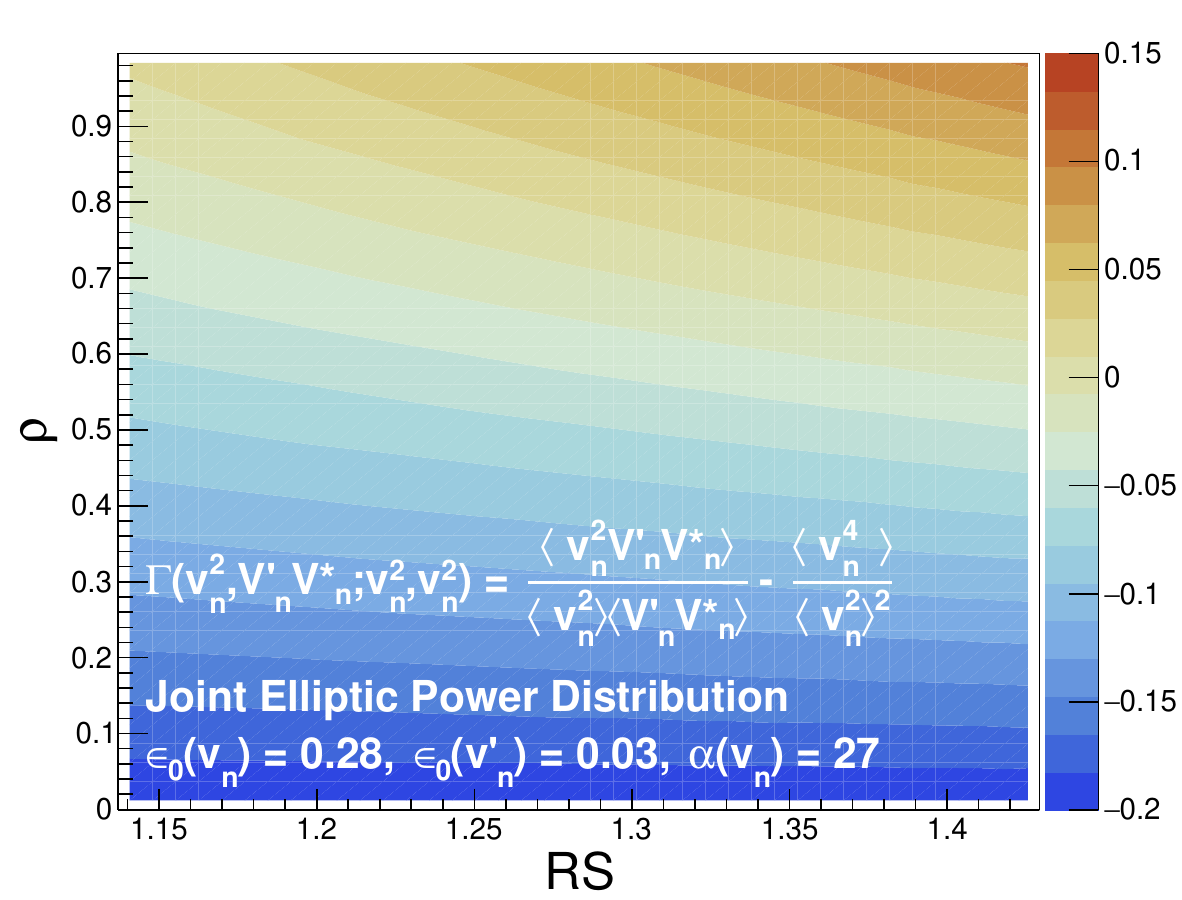} & 
 \includegraphics[width=9cm]{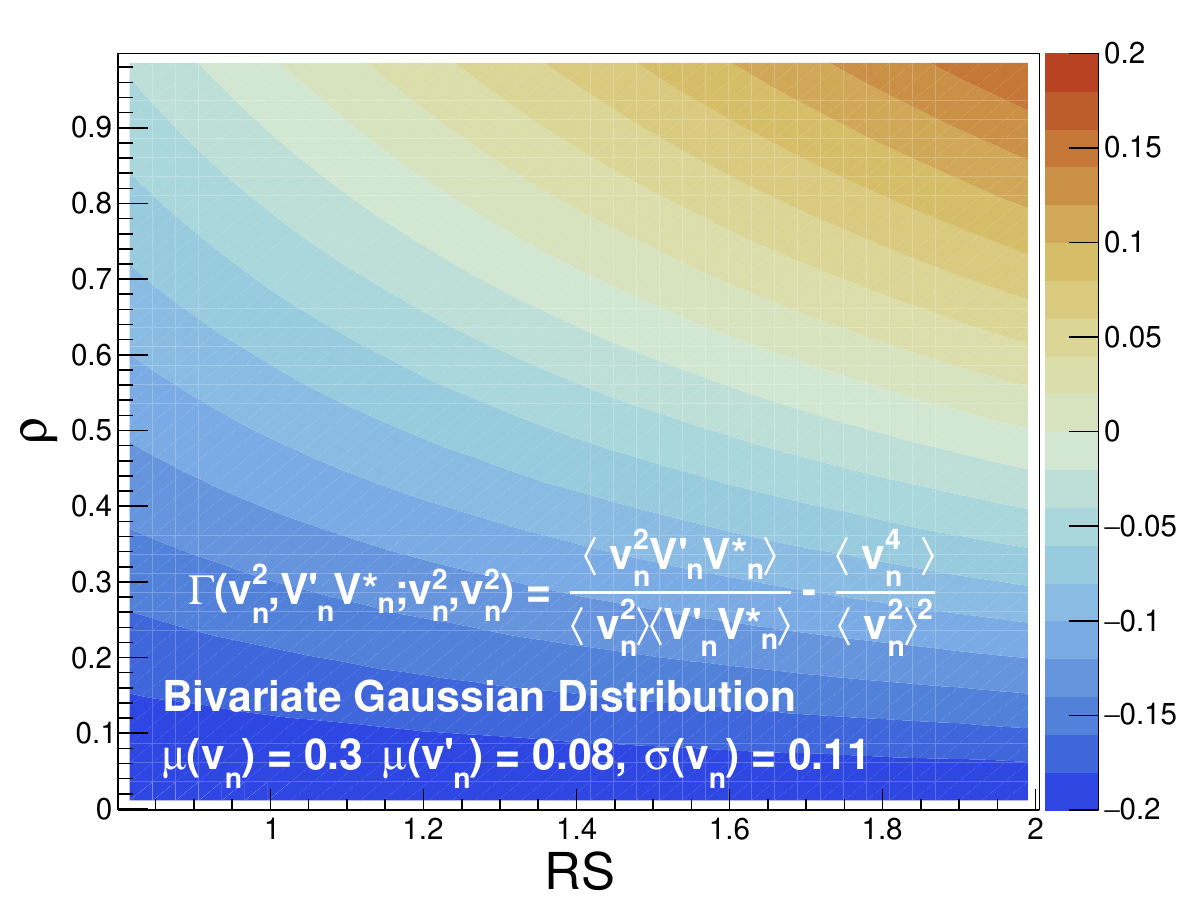}
 \end{tabular}
 \caption{The sensitivity of $\Gamma(v_n^2,V'_nV_n^{*};v_n^2,v_n^2)$ is shown as a function of $\rho(v_n,v'_n)$ and RS $= \sigma(v'_n)\mu(v_n)/\sigma(v_n)\mu(v'_n)$ for $v'_n$ and $v_n$ parametrized using the joint elliptic power distribution (left), and a bivariate Gaussian distribution (right). Note that the range of RS is different in the two plots.}
 \label{fig:deltash}
\end{figure}

$\Gamma(v_n^2,V'_nV_n^{*};v_n^2,v_n^2)$ was originally proposed in Ref. \cite{Betz:2016ayq} (referred to there as $\Delta_n^{sh}$) with the intent of evaluating a difference in fluctuations between $v'_n$ and $v_n$. Unfortunately, it does not suffice in this goal on its own, as it demonstrates sensitivity to both $\rho$ and to RS, which is apparent for both of the Gaussian and elliptic power parametrizations seen in Fig. \ref{fig:deltash}. However, it does still have significant standalone importance. The authors of Ref. \cite{Betz:2016ayq} produced predictions for this quantity using a jet quenching hydrodynamical model, which can be used to constrain both $\rho$ and RS.

 Fig.~\ref{fig:deltash} shows explicitly the sensitivity of $\Gamma(v_n^2,V'_nV_n^{*};v_n^2,v_n^2)$ as defined below,
\begin{equation}\label{eq:Gamma}\Gamma(v_n^2,V'_nV_n;v_n^2,v_n^2) = \frac{\avg{v_n^2V'_nV_n^{*}}}{\avg{V'_nV_n}\avg{v_n^2}} - \frac{\avg{v_n^4}}{\avg{v_n^2}^2} = nSC(V'_nV_n,v_n^2) - nSC(v_n^2,v_n^2)\end{equation}
to fluctuations in $\rho$ and RS. Since $\Gamma(v_n^2,V'_nV_n^*;v_n^2,v_n^2)$ can be written as a difference of symmetric cumulants, we can see that positive and negative values for $\Gamma$ show whether $cov(V'_nV_n^*,v_n^2)$ is larger or smaller than $\sigma^2(v_n^2)$. In this way, $\Gamma$ is a measurement of the magnitude of fluctuations in $V'_nV_n^*$ relative to $v_n^2$.
$\Gamma$ also has considerable importance because it only has a first order dependence on $v'_n$, making it more feasible to measure with jets or other rare probes, as it has the same statistical requirements as the correlations used to create differential cumulants \cite{Borghini:2001vi}. 

The general behavior of $\Gamma(v_n^2,V'_nV_n^{*};v_n^2,v_n^2)$ is desirable; it is largely monotonic in both $\rho$ and RS, as shown in Fig. \ref{fig:deltash}, and demonstrates comparable values for both parametrizations. While there is a clear difference between the right and left panels, we see that the values and behavior of the observable are similar when considering that the RS for the elliptic power distribution varies over a much smaller range of values compared to RS for the bivariate Gaussian. The level of correspondence between both $p(v_n)$ parametrizations is discussed in depth in Sec.~\ref{sec:5}. 

\subsubsection{$\zeta(V'_nV_n^{*},V'_nV_n^{*};v_n^2,v_n^2)$}

 The quantity $\zeta(V'_nV_n^{*},V'_nV_n^{*};v_n^2,v_n^2)$ can be understood as the ratio of the non-constant terms in two normalized symmetric cumulants. While $\zeta$ can not in general express the differences between central moments, it describes the relative magnitude of the fluctuations between $V'_nV_n^*$ and $v_n^2$, as detailed below,
\begin{equation}
 \zeta(V'_nV_n^{*},V'_nV_n^{*};v_n^2,v_n^2) = \frac{\avg{(V'_nV_n^{*})^2}}{\avg{V'_nV_n^{*}}^2} \bigg/ \frac{\avg{v_n^4}}{\avg{v_n^2}^2}
\end{equation}
 where a higher order dependence on $v'_n$ can be seen for this quantity, allowing for a different sensitivity to both RS and $\rho$.

\begin{figure}
 \centering
 \begin{tabular}{ll}
 \includegraphics[width=9cm]{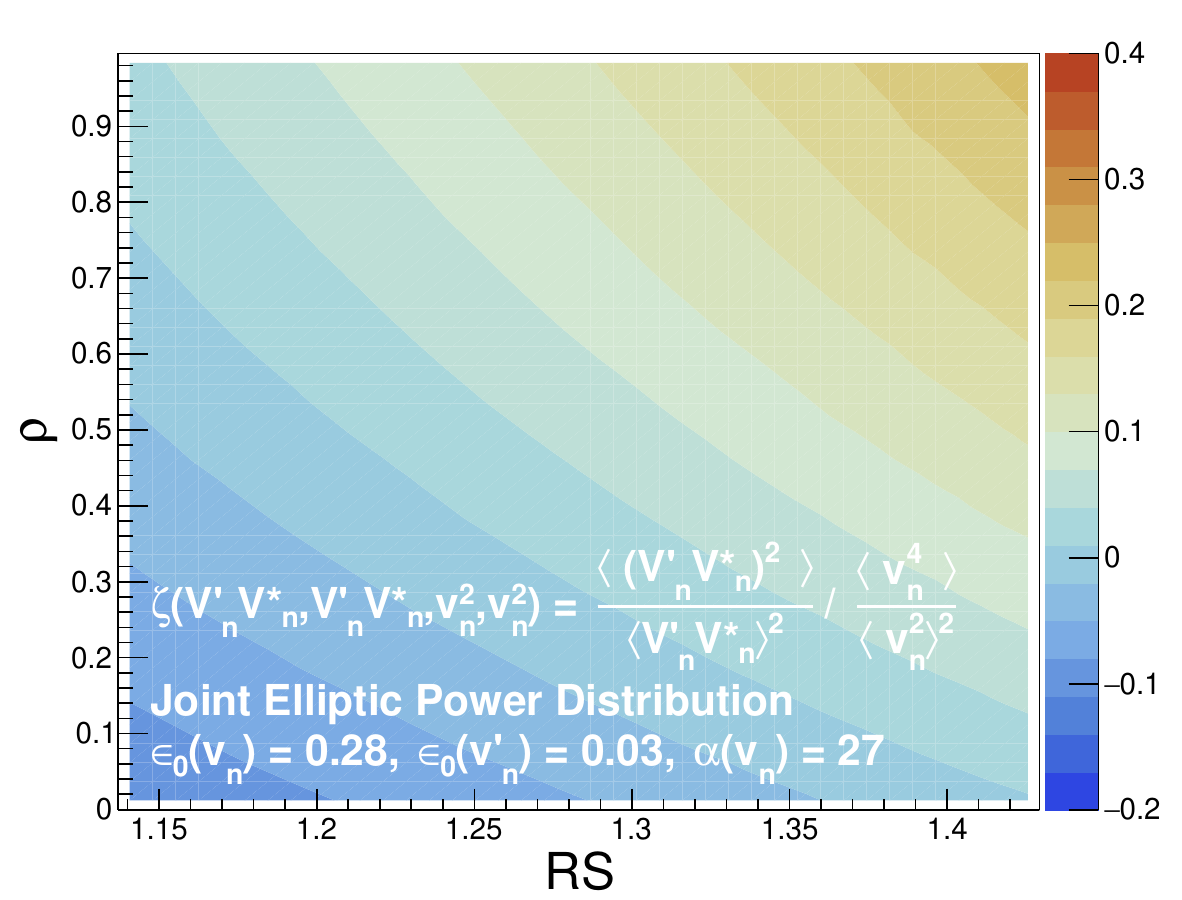} & 
 \includegraphics[width=9cm]{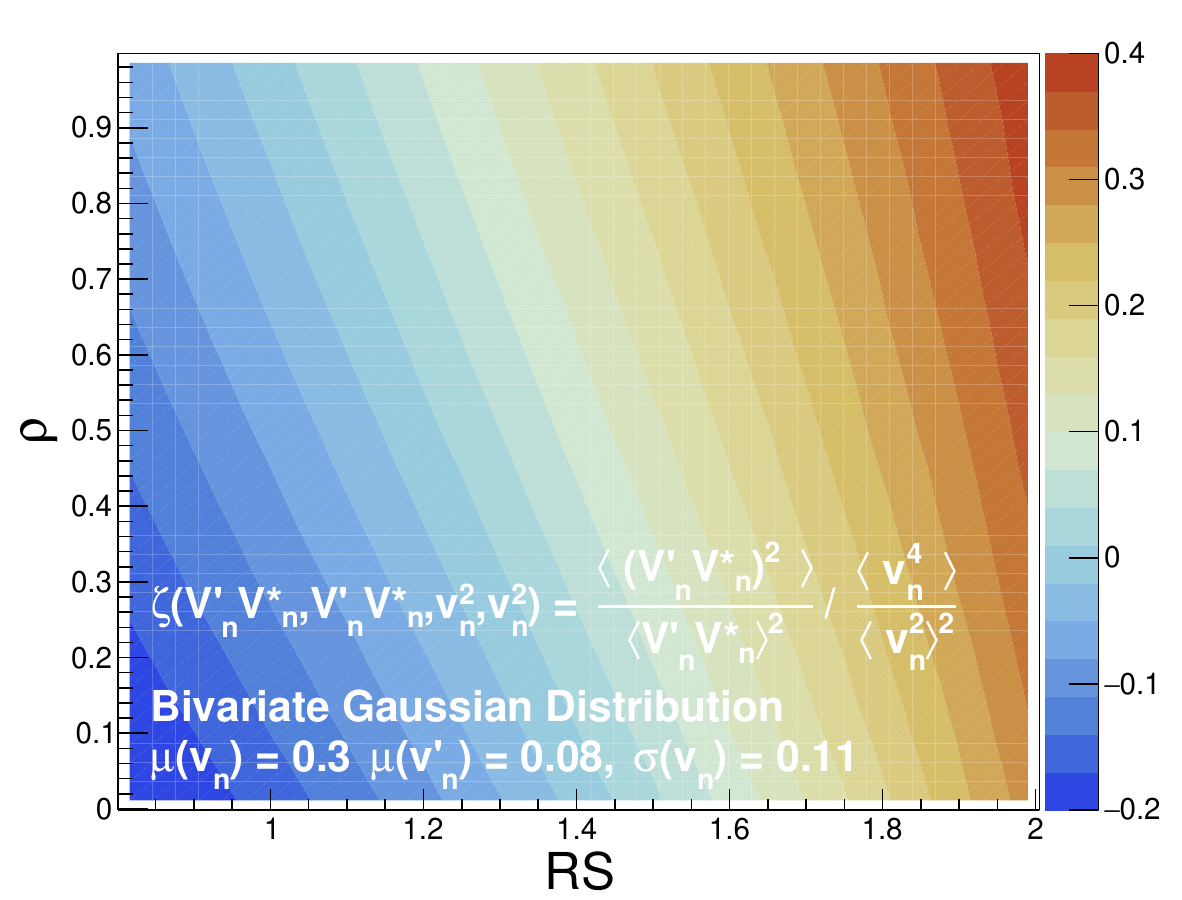}
 \end{tabular}
 \caption{The sensitivity of $\zeta(V'_nV_n^{*},V'_nV_n^{*};v_n^2,v_n^2)$ is shown as a function of $\rho(v_n,v'_n)$ and RS for $v'_n$ and $v_n$ parametrized using the joint elliptic power distribution (left), and a bivariate Gaussian distribution (right). Note that the range of RS is different in the two plots.}
 \label{fig:zeta}
\end{figure}
 Fig. \ref{fig:zeta} shows the sensitivities displayed by $\zeta$ to both $\rho$ and RS. We find that $\Gamma$ strongly varies with RS, but is less sensitive to $\rho$. This behavior is quite different compared to $\Gamma$, which increases more with $\rho$ and is nearly flat in RS. Additionally, the range of values covered by $\zeta$ is larger than that of $\Gamma$. This comes from a comparatively stronger second order dependence on $v'_n$, as well as from $\zeta$ expressing a ratio rather than a difference in fluctuations. Fig. \ref{fig:zeta} demonstrates that $\zeta$, like $\Gamma$, monotonically increases along with both RS and $\rho$, but with very different sensitivities. The difference in sensitivities between $\Gamma$ and $\zeta$ ensures that there is little redundancy in simultaneous measurements, and, as described in Sec.~\ref{sec:5}, may assist in the discernment of values for both $\rho$ and RS. 

\subsubsection{nSC$({v'}_n^2,v_n^2)$}

\begin{figure}
 \centering
 \begin{tabular}{ll}
 \includegraphics[width=9cm]{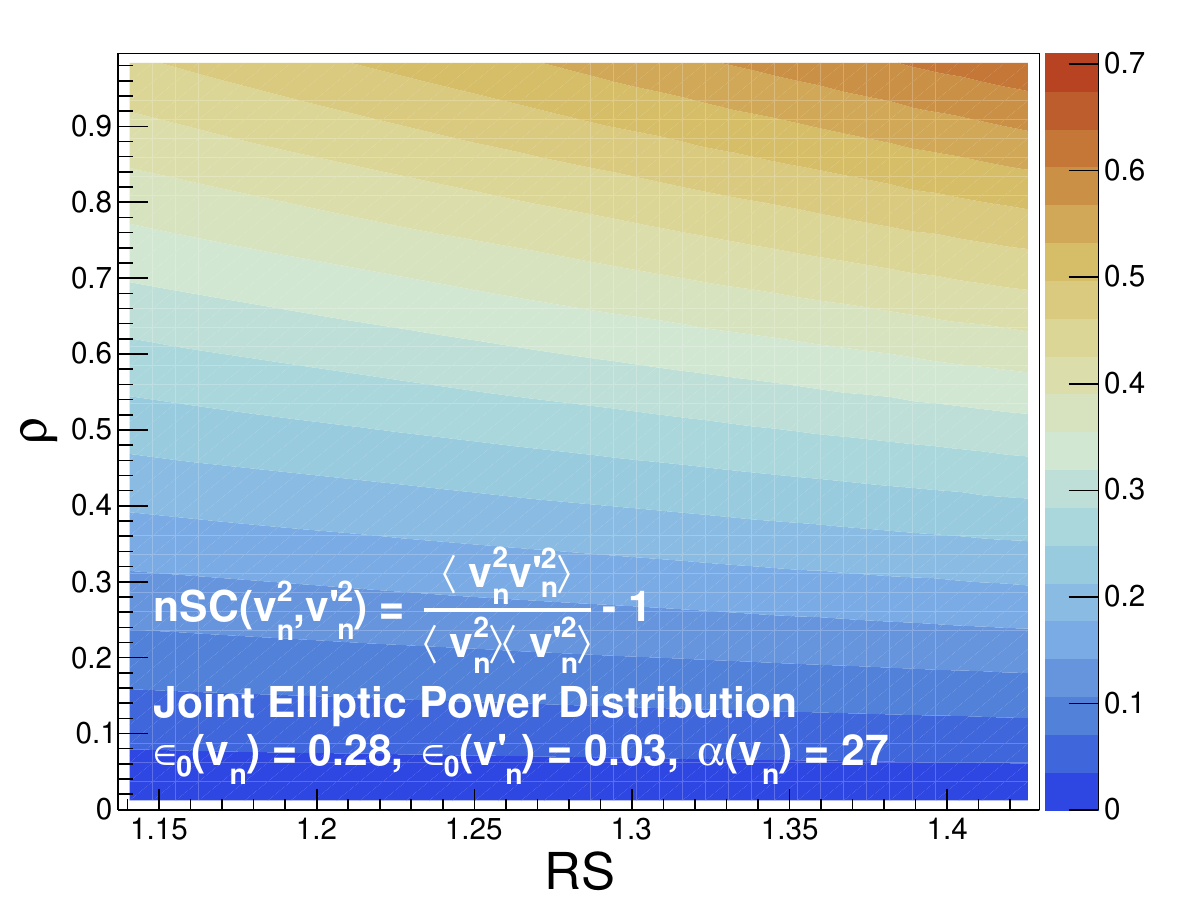} & 
 \includegraphics[width=9cm]{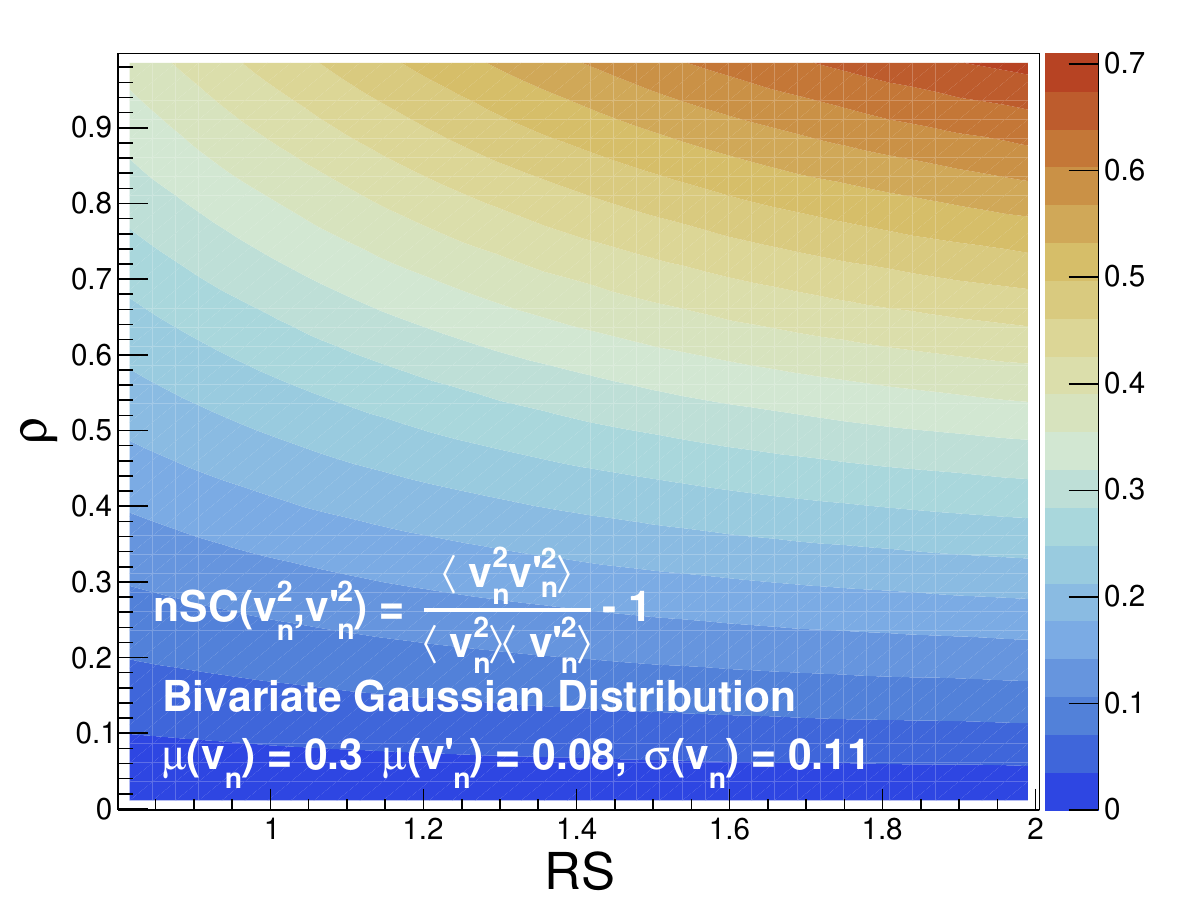}
 \end{tabular}
 \caption{The sensitivity of nSC$({v'}_n^2,v_n^2)$ is shown as a function of $\rho(v_n,v'_n)$ and RS for $v'_n$ and $v_n$ parametrized using the joint elliptic power distribution (left), and a bivariate Gaussian distribution (right). Note that the range of RS is different in the two plots.}
 \label{fig:scz}
\end{figure}

 Symmetric cumulants have been used to probe flow coefficient magnitude correlations \cite{Bilandzic:2013kga,Giacalone:2016afq,ATLAS:2018ngv}, and their interpretation is very straightforward: larger values indicate a stronger correlation between ${v'}_n^2$ and $v_n^2$. Since covariance is a co-central moment, we can also interpret that the value of nSC$({v'}_n^2,v_n^2)$ indicates the correlation between a departure from the mean in ${v'}_n^2$ with a departure from the mean in $v_n^2$.
Fig.~\ref{fig:scz} shows the sensitivity of a normalized symmetric cumulant nSC$({v'}_n^2,v_n^2)$ measuring a normalized covariance between ${v'}_n^2$ and $v_n^2$:
\begin{equation}
 nSC({v'}_n^2,v_n^2) = \frac{\avg{{v'}_n^2{v}_n^2}}{\avg{{v'}_n^2}\avg{{v}_n^2}}-1 = \frac{\avg{\left({v'}_n^2-\avg{{v'}_n^2}\right)\left(v_n^2-\avg{v_n^2}\right)}}{\avg{{v'}_n^2}\avg{v_n^2}}.
\end{equation}

 The sensitivities of the observable nSC$({v'}_n^2,v_n^2)$ to $\rho$ and RS, are shown in Fig.~\ref{fig:scz}. The observable nSC$({v'}_n^2,v_n^2)$ is a function of the stochastic variable ${v'}_n^2$ in contrast to the variable $V'_nV_n^*$ which was used for the past two observables. Like the other observables, nSC$({v'}_n^2,v_n^2)$ is monotonically increasing in both RS and $\rho$. Interestingly, the dependence of nSC$({v'}_n^2,v_n^2)$ is qualitatively very similar to $\Gamma$, which requires a smaller sample of POIs to measure accurately. However, nSC displays a larger range of values, and has a fundamentally different interpretation, owing to a dependence on different random variables than $\Gamma$. Finally, when considering the restricted domain for the joint elliptic power distribution, we can see that the observable yields very similar values across both $\rho$ and RS.

\subsubsection{nASC$({v'}_n^2,v_n^2,v_n^2)$}

\begin{figure}
 \centering
 \begin{tabular}{ll}
 \includegraphics[width=9cm]{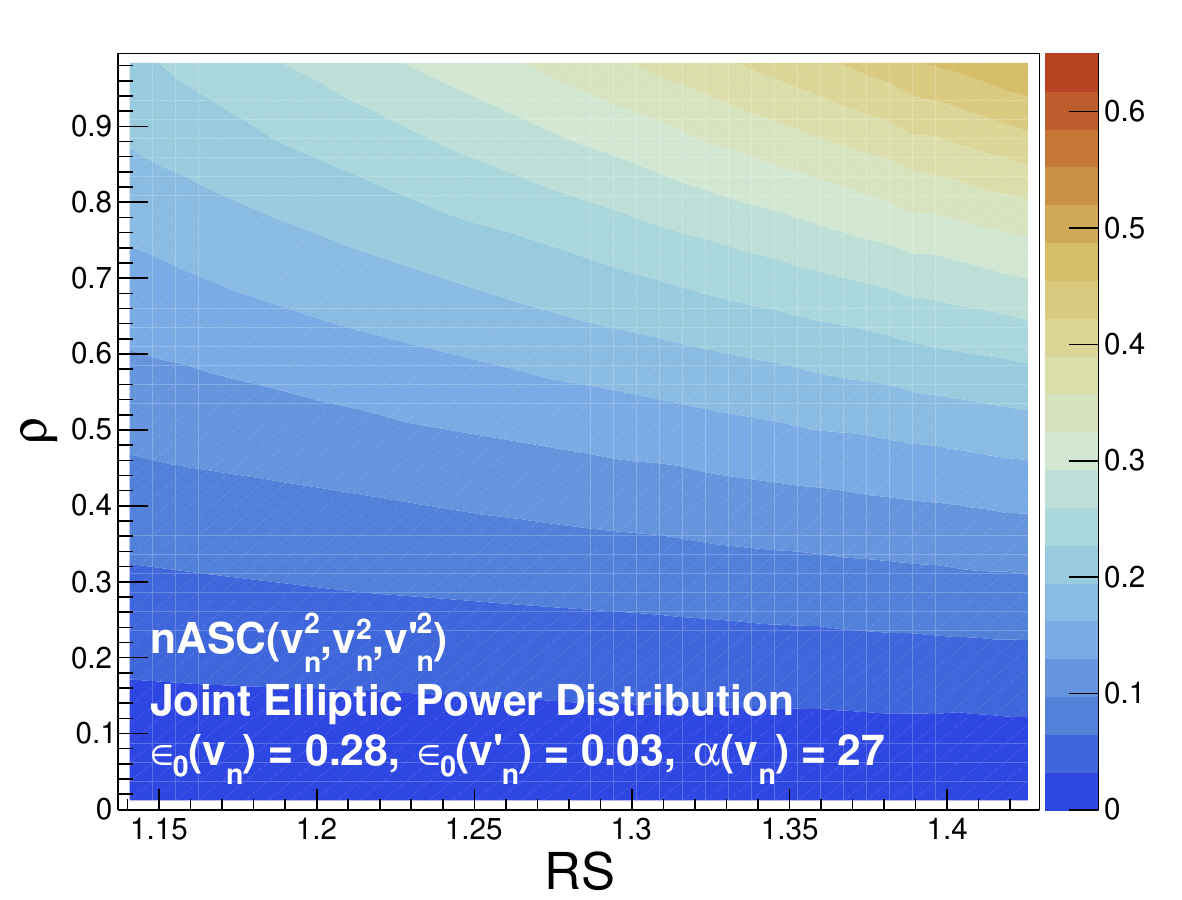} & 
 \includegraphics[width=9cm]{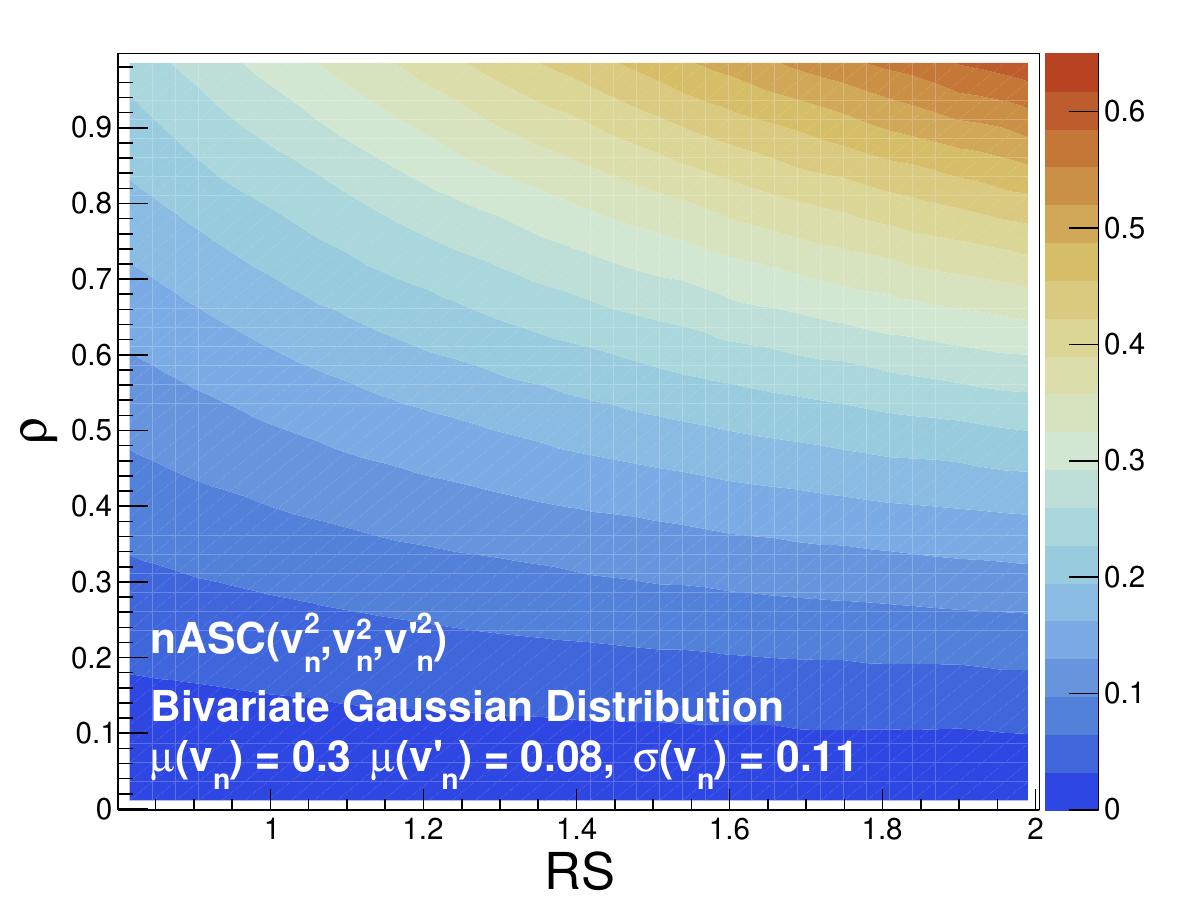}
 \end{tabular}
 \caption{The sensitivity of nASC$({v'}_n^2,v_n^2,v_n^2)$ is shown as a function of $\rho(v_n,v'_n)$ and RS for $v'_n$ and $v_n$ parametrized using the joint elliptic power distribution (left), and a bivariate Gaussian distribution (right). Note that the range of RS is different in the two plots.}
 \label{fig:nASCs}
\end{figure}

The sensitivities of the normalized asymmetric cumulant nASC$({v'}_n^2,v_n^2,v_n^2)$ are shown in Fig.~\ref{fig:nASCs}, and have been used to describe higher order fluctuations between flow harmonics \cite{Bilandzic:2021rgb,ALICE:2021klf,ATLAS:2018ngv}. Asymmetric cumulants have also been generalized to incorporate multi-differential correlations in Ref.~\cite{Holtermann:2023vwr}, and provide a measurement of the genuine correlations between larger sets of stochastic variables. The observable is included here to illustrate a higher-order generalization of nSC$({v'}_n^2,v_n^2)$. The asymmetric cumulant we study, nASC$({v'}_n^2,v_n^2,v_n^2)$, measures fluctuations between ${v'_n}^2$ and $v_n^2$, displaying second order dependence on $v_n^2$:

\begin{equation}\label{nasc}
 nASC({v'}_n^2,v_n^2,v_n^2) = \frac{\avg{{v'}_n^2v_n^4} - \avg{{v'}_n^2}\avg{v_n^4} - 2\avg{v_n^2}\avg{v_n^2{v'}_n^2} + 2\avg{{v'}_n^2}\avg{v_n^2}^2}{\avg{{v'}_n^2}\avg{v_n^2}^2} = \frac{\avg{\left({v'_n}^2 - \avg{{v'_n}^2}\right)\left({v_n}^2 - \avg{{v_n}^2}\right)^2}}{{\avg{{v'}_n^2}\avg{v_n^2}^2}}.
\end{equation}
The sensitivities of nASC$({v'}_n^2,v_n^2,v_n^2)$ to RS and $\rho$ in our toy model can be seen in Fig.~\ref{fig:nASCs}. nASC clearly increases along the horizontal and vertical axes, indicating significant sensitivity to both $\rho$ and RS. The dependence of nASC on both RS and $\rho$ appears to increase more non-linearly compared to nSC and $\Gamma$, but overall has a similar qualitative appearance, meaning that in this context, a measurement of nASC may not add significant information to that provided by a measurement of $\Gamma$ or nSC. Recalling that this asymmetric cumulant is a central moment, we can understand the positive result seen at high values for RS as indicative of a very strong relationship between deviations of ${v'}_n^2$ from the mean, and the overall variance in the distribution of $v_n^2$, $\avg{(v_n^2 - \avg{v_n^2})^2}.$

\section{Applicability}\label{sec:5}

Now that we have discussed the sensitivities of four different correlation observables using our toy model, we explain how this phenomenological work can be used to extract values for $\rho$ and RS from experimental data. We then address considerations about the choices of parametrizations and correlation models used throughout this paper.

\subsection{Constraining RS and $\rho$}

Given a measurement of any observable detailed in this paper and sufficient inputs for the toy model specified in Sec.~\ref{sec:4} (bivariate azimuthal anisotropy model $p(v'_n,v_n)$ and a set of fixed parameters), one would be able to constrain the values of RS and $\rho$ to a curve in RS, $\rho$ space. Depending on the observables and their sensitivity, this alone may be a substantial constraint, especially in the context of theoretical predictions.
\begin{figure}
 \centering
 \includegraphics[width = 15cm]{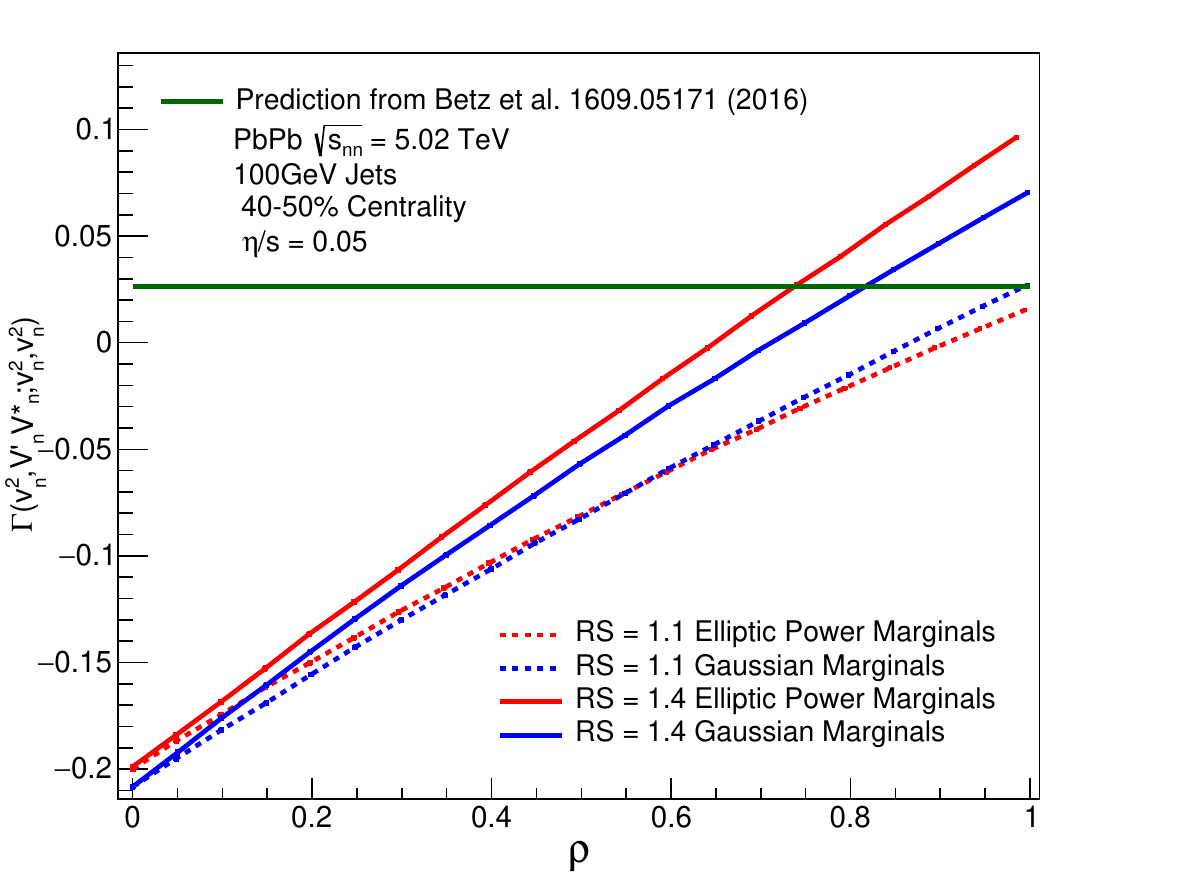}
 \caption{Values for $\Gamma(V'_nV_n^*,v_n^2;v_n^2,v_n^2)$ are displayed as a function of $\rho$, for both elliptic power (red) and Gaussian (blue) marginal distributions, and at RS = 1.1 (dotted lines) and RS = 1.4 (solid lines), alongside a prediction for $\Gamma(V'_nV_n^*,v_n^2;v_n^2,v_n^2)$ found in Ref. \cite{Betz:2016ayq} for PbPb collisions at $\sqrt{s_{N N}}=5.02$ TeV, with $\eta/s = 0.05$, at $40-50\%$ centrality for jets at $p_T =$ 100 GeV.}
 \label{fig:prediction}
\end{figure}
As an example, in Fig. \ref{fig:prediction}, we plot values for $\Gamma(V'_nV_n^*,v_n^2;v_n^2,v_n^2)$ both with elliptic power and Gaussian marginals, as a function of $\rho$ for two different RS values, RS = $1.1$ and RS = $1.4$. These RS values were selected because they cover most of the range of RS for the joint elliptic power distribution, shown in Fig.~\ref{fig:alphav}. Predictions for this quantity using a combined jet-hydrodynamical model to predict energy loss were published in Ref. \cite{Betz:2016ayq}, for PbPb collisions at $\sqrt{s_{N N}}=5.02$ TeV, measuring $\Gamma(V'_nV_n^*,v_n^2;v_n^2,v_n^2)$ for $\eta/s = 0.05$, using jets at $p_T =$ 100 GeV. We show the value obtained in Ref. \cite{Betz:2016ayq} for $40-50\%$ centrality, using the elliptic power distribution parameters from Fig.~\ref{fig:glauberfits}. 
The predicted value from Ref.~\cite{Betz:2016ayq} intersects computed values of $\Gamma(V'_nV_n^*,v_n^2;v_n^2,v_n^2)$ at $\rho \geq 0.6$, indicating that for $v_n$ and $v'_n$ modeled using the joint elliptic power distribution with fixed values detailed in Table~\ref{tab:params}, $\rho \geq 0.6$. While this relationship differs at different centralities, it suggests that experimental models predict relatively large correlations between $v'_n$ and $v_n$ even for jets at high transverse momentum. It is worth noting that this relationship persists for RS significantly greater than 1, where despite having mean values and fluctuations of different magnitudes, $v_n$ and $v'_n$ still remain strongly correlated. 

Owing to the unique sensitivities displayed by each observable to RS and $\rho$ as shown in Figs.~\ref{fig:deltash}-\ref{fig:nASCs}, for any \textit{two or more} observables that display monotonicity for the inputs, we can obtain RS and $\rho$ values by determining where the level curves (curves in RS, $\rho$ space where an observable has a constant value) for each observable intersect. Using plots such as Fig. \ref{fig:nASCs}, alongside experimental measurements of a given observable, we can identify the level curves in RS, $\rho$ space occupied by each observable. Combining measurements of multiple observables with different sensitivities allows for the confinement of the possible values for RS and $\rho$ to separate curves in RS,$\rho$ space, each corresponding to the level curve of a measured observable. This means that the actual values for RS and $\rho$ can be determined by locating where all of the level curves intersect. For a given parametrization, only two level curves are necessary, but to check for consistency, or to distinguish a larger number of fluctuation quantities (e.g. RS, $\rho$, and $\sigma(v_n)$), any number of simultaneous measurements can be used. 
The uncertainties measured for any observable detailed in this paper will likewise constrain $RS$ and $\rho$ to bands in RS,$\rho$ space containing all level curves within the uncertainties measured for the observable. The intersection of these bands will provide a two dimensional overlap region in RS,$\rho$ space covering all possible points for which both observables take values within their respective uncertainties. Measuring three or more observables simultaneously will force the overlap region of uncertainty bands for each observable to be further reduced, decreasing the region in which all observables remain within their uncertainties, and providing a stronger constraint on RS and $\rho$.

\begin{figure}
 \centering
 \includegraphics[width = 12cm]{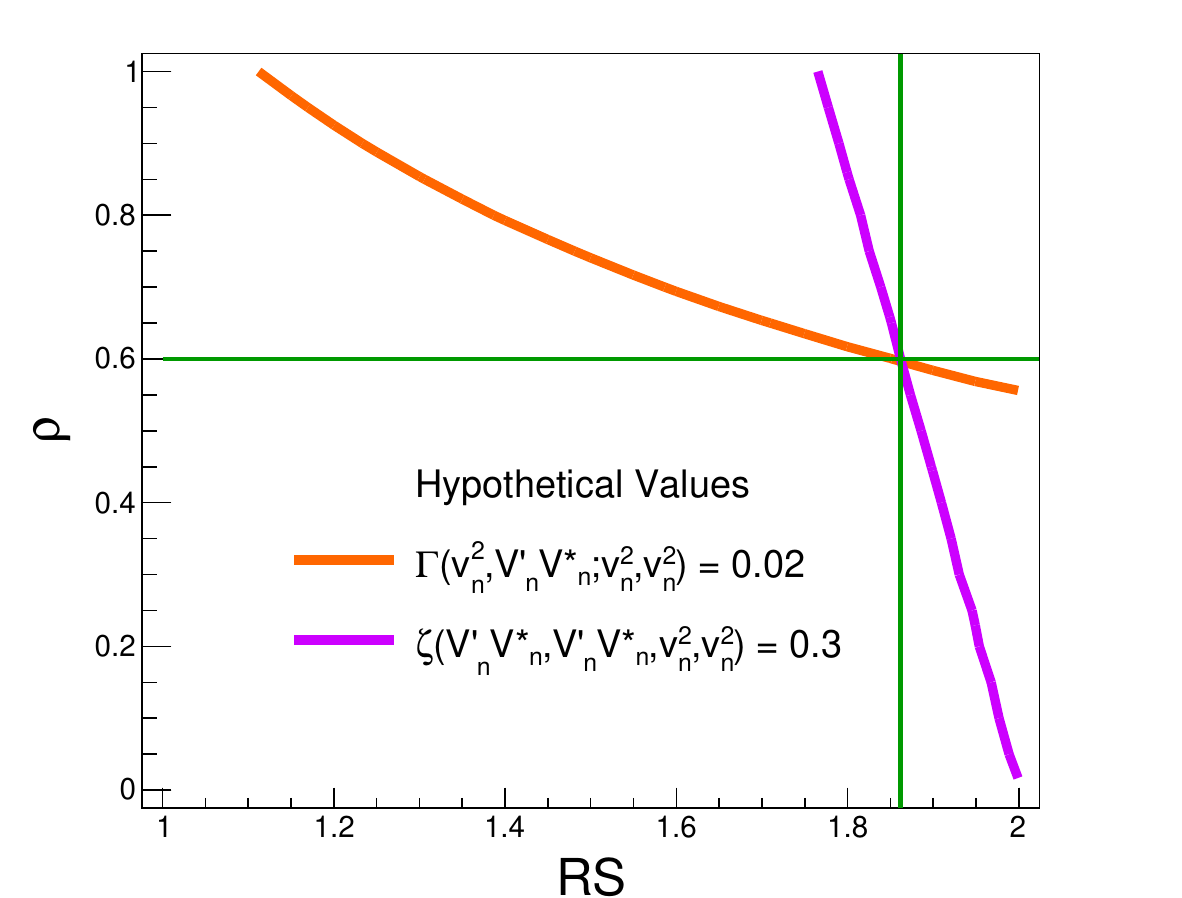}
 \caption{The level curves for $\Gamma(v_n^2,V'_nV^*_n;v_n^2,v_n^2) = 0.02$ (orange) and $\zeta(V'_nV_n^*,V'_nV_n^*,v_n^2,v_n^2) = 0.3$ (purple) for varying values of $\rho$ and RS for the bivariate Gaussian distribution are shown. The values for RS and $\rho$ where the level curves intersect are also labeled (green).}
 \label{fig:hyp}
\end{figure}

A hypothetical example of the intersection of level curves for nASC and $\Gamma$ is shown in Fig. \ref{fig:hyp}. In this figure, the level curves corresponding to hypothetical measurements of $\Gamma(v_n^2,V'_nV^*_n;v_n^2,v_n^2) = 0.02$ and $\zeta(V'_nV_n^*,V'_nV_n^*,v_n^2,v_n^2) = 0.3$ are shown. Their intersection indicates that in this toy model, RS $ \approx 1.85$ and $\rho$ $\approx 0.6$ is the only point in RS, $\rho$ space satisfying $\Gamma(v_n^2,V'_nV^*_n;v_n^2,v_n^2) = 0.02$ and $\zeta(V'_nV_n^*,V'_nV_n^*,v_n^2,v_n^2) = 0.3$.
In general, simply measuring a four-particle correlation with two POI enables the measurement of enough unique observables, (with different normalizations, stochastic variables, construction, and sensitivity) such as those detailed in Sec.~\ref{observablesection}, to make the analysis of two or more intersecting level curves possible.

The toy model detailed in this paper uses joint distributions, $p(v_n,v'_n),$ parametrized with five variables, as seen in Table~\ref{tab:params}. For each parametrization, only two parameters are varied to evaluate sensitivities to RS and $\rho$. These two variables were selected due to their ability to characterize fluctuations in $p(v'_n)$, and correlations between $v_n$ and $v'_n$. Additionally, most existing azimuthal anisotropy measurement techniques and observables (such as the scalar product, or reference multiparticle cumulants) are not sensitive to either RS or $\rho$. However, in each parametrization, the remaining fixed parameters can also be important to the determination of an observable's sensitivity to both RS and $\rho$. In Appendix~\ref{ap:Calc}, we detail a number of nonlinear relationships between the parameter values and multivariate moments for a bivariate Gaussian distribution $p(v_n,v'_n)$, illustrating how even the fixed parameters $\mu(v_n), \mu(v'_n),$ and $\sigma(v_n)$, can have significant effects on the observables described in this paper.

The fixed parameters in this model, for both the joint elliptic power and bivariate Gaussian distributions, characterize $p(v_n)$ and $\avg{v'_n}$. However, there may exist quantities that drive changes in the parameters defining $p(v_n)$ and $\avg{v'_n}$. An example of such a quantity is centrality, as $p(v_n),$ and parameters $\sigma(v_n)$ and $\mu(v_n)$ take on significantly different values for events with differing centrality \cite{ATLAS:2018ngv}. Another example of such a quantity is the harmonic $n$. The shape of the probability distribution $p(v_n)$ changes significantly between $v_2,v_3,$ and $v_4$ as documented in Ref.~\cite{ATLAS:2013xzf}. To accurately apply this toy model in any analysis, the "fixed parameters" must be reevaluated for variations in every quantity that drives changes in $p(v_n)$ or $\avg{v'_n}$. As an example, parameters characterizing $p(v_n)$ should be evaluated separately for each centrality bin and for each harmonic before reapplying the toy model.

\subsection{Additional Considerations}

A bivariate parametrization $p(v_n,v'_n)$ is necessary to generate data for the toy model defined in this paper. In this section, we address the effects of different choices for both the marginal distributions, $p(v_n)$ and $p(v'_n)$, and the correlation model used to generalize $p(v_n),p(v'_n)$ into a bivariate distribution. 

While the Gaussian distribution represents Glauber simulation results less accurately compared to the elliptic power distribution as seen in Fig.~\ref{fig:glauberfits}, we have found the Gaussian distribution to be more comprehensive as a model for both $p(v_n)$ and $p(v'_n)$. Using the Gaussian distribution, fluctuations in $p(v_n)$ and $p(v'_n)$ are easily characterized and varied, and the marginal distributions $p(v_n)$ and $p(v'_n)$ generalize naturally with the Gaussian copula model into a bivariate Gaussian distribution.
The elliptic power distribution, more accurately represents the eccentricity distributions generated by the Glauber model \cite{Yan:2014afa} as shown in Fig. \ref{fig:glauberfits}. However, it demonstrates a limited range of RS for realistic parameter values, and demonstrates qualitative consistency with sensitivities obtained from the bivariate Gaussian distributions. Figures \ref{fig:deltash} - \ref{fig:nASCs} indicate that for each observable, the values from the bivariate Gaussian and joint elliptic power distributions are qualitatively similar, and display similar sensitivities to both RS and $\rho$. 

\begin{figure}
\centering
\begin{tabular}{ll}
\includegraphics[width=9cm]{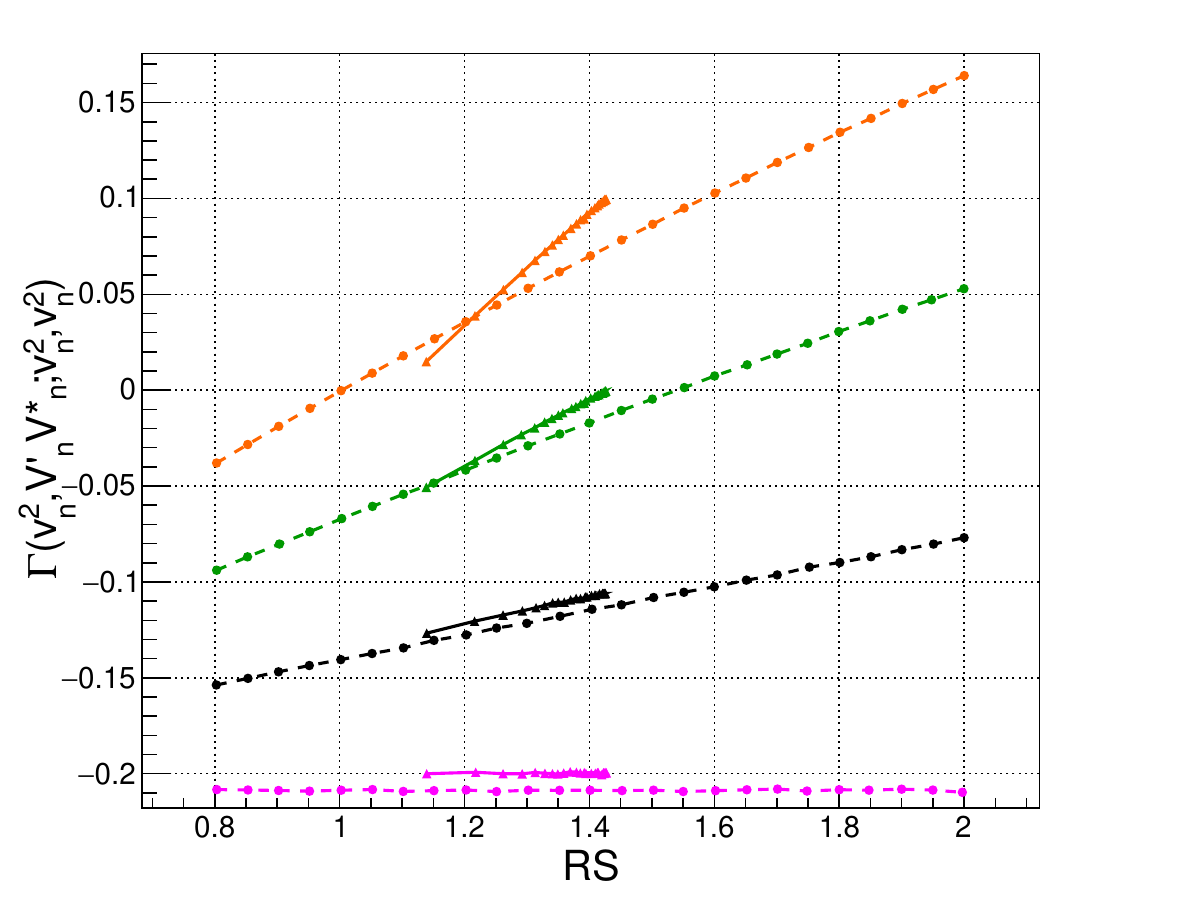} &
\includegraphics[width=9cm]{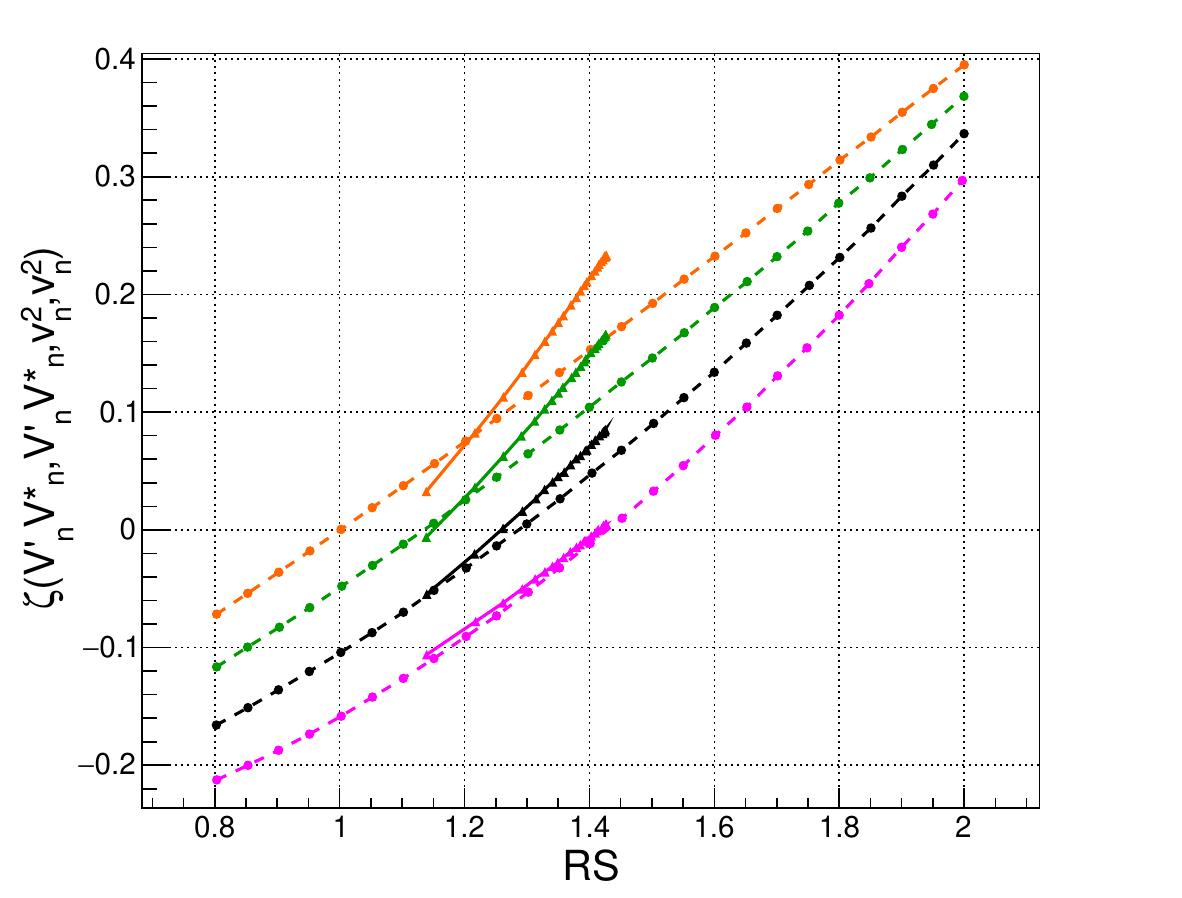}\\
\includegraphics[width=9cm]{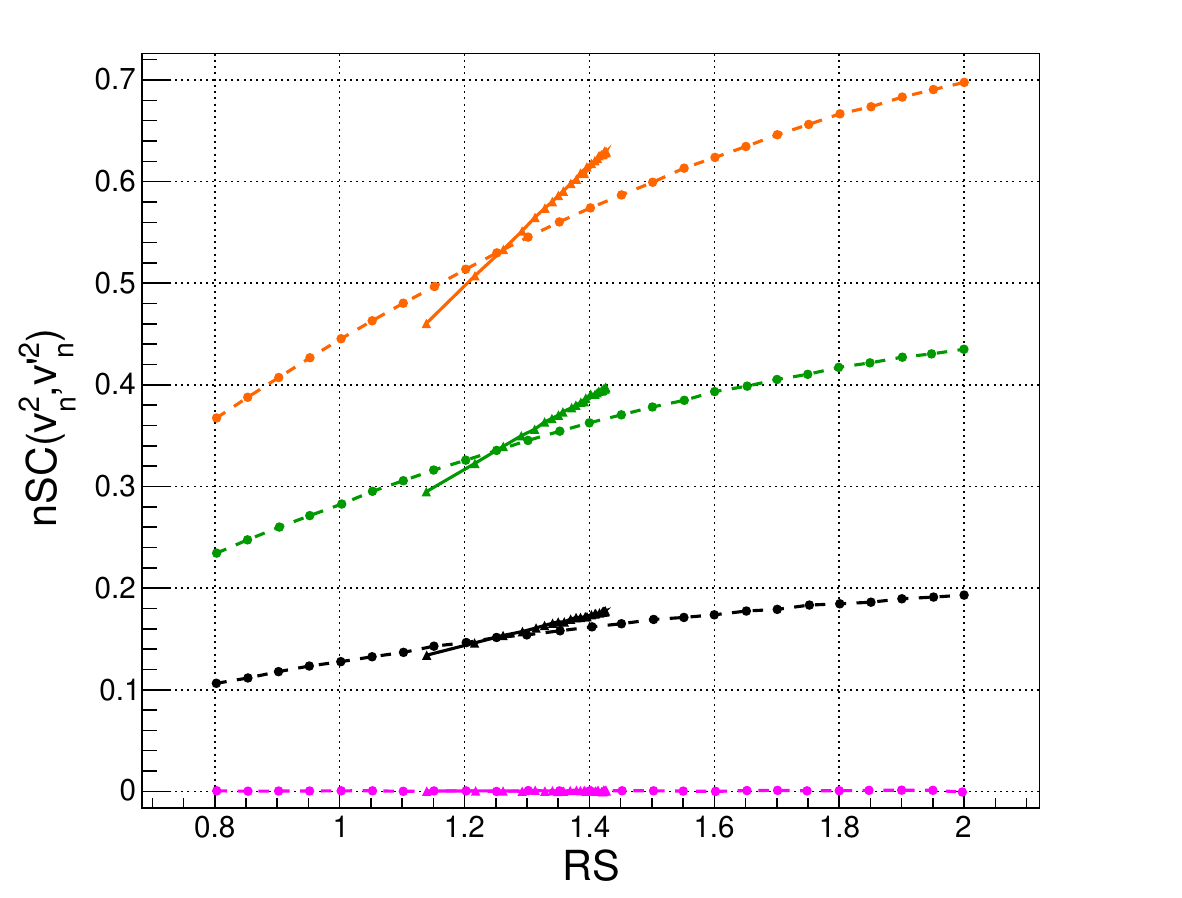}& 
\includegraphics[width=9cm]{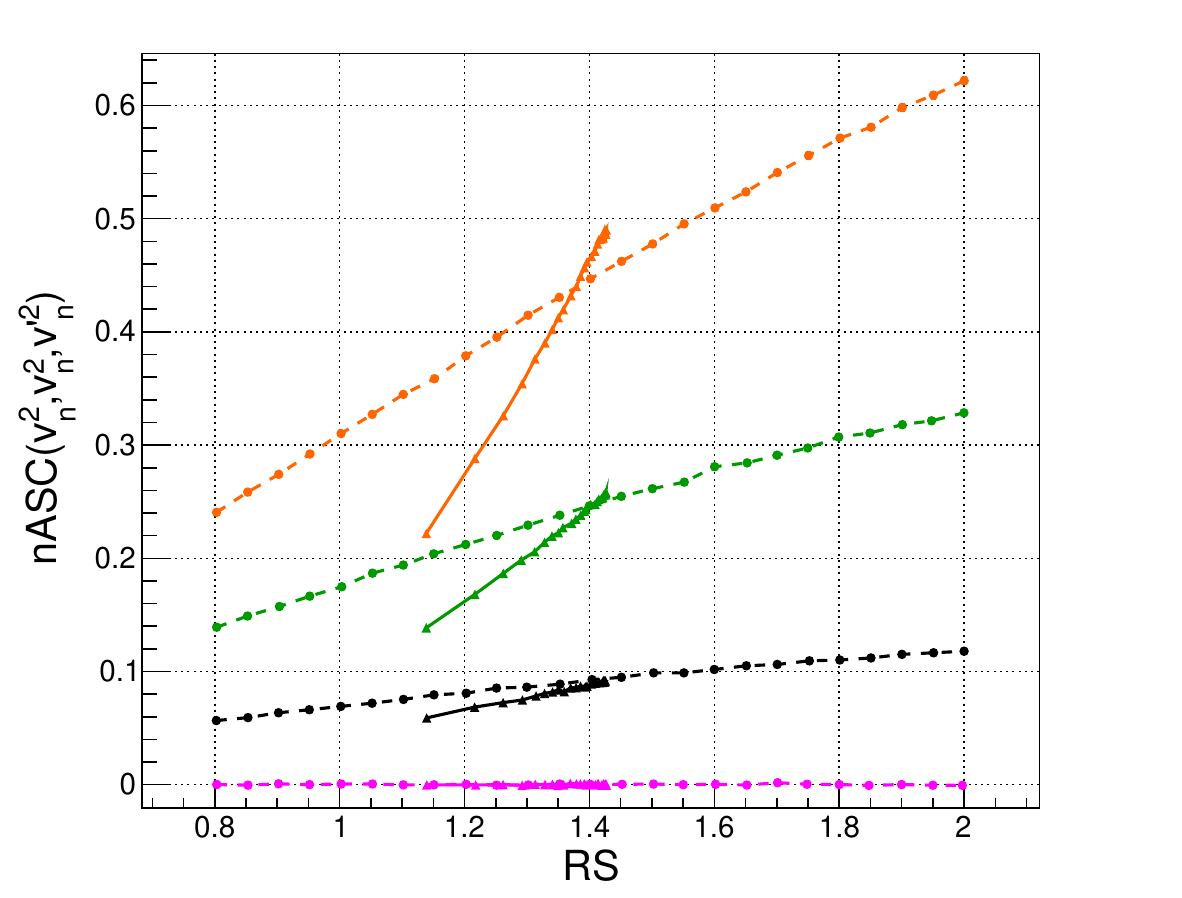}\\
\end{tabular}
\includegraphics[width = 9cm]{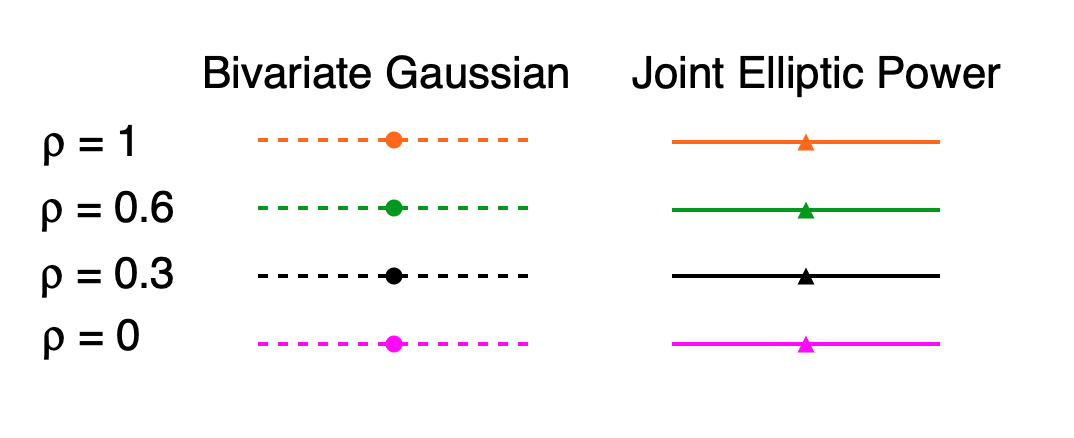}
\caption{The values for $\Gamma$ (top left), $\zeta$ (top right),nSC (bottom left), and nASC (bottom right) are plotted as functions of the relative spread for selected values of $\rho$, for both elliptic power marginal and Gaussian marginal distributions.}
\label{fig:grid}
\end{figure}

To better assess the sensitivity of each observable to the choice of parametrization, we provide a direct comparison. In Fig.~\ref{fig:grid}, we display the values of each observable discussed in Sec. \ref{subsec:4} for both the joint elliptic power distribution and the bivariate Gaussian distribution as a function of RS for four selected values of $\rho$. In each plot, the correspondence between the values of each observable derived from the joint elliptic power distribution and the bivariate Gaussian distribution can be assessed. 

The observables nSC, $\zeta$, and $\Gamma$ demonstrate differences in value between both parametrizations, but this difference is far greater for nASC. This is unsurprising because nASC is constructed using six-particle correlations, and thus has sensitivity to a joint moment of order six. This sixth order moment is more sensitive to the differing tail behavior between the elliptic power and Gaussian parametrizations than any moment used in the construction of nSC$,\zeta$ or $\Gamma$, since they only rely on moments of order at most four.
Additionally, Fig.~\ref{fig:grid} shows that the clearest differences between our two parameterizations show up at high $\rho$, whereas our two distributions provide nearly identical results at low $\rho$ values. This is because highly correlated bivariate distributions $p(v_n,v'_n)$ are constrained more narrowly onto a line within their domain. As a result, their probability densities become more compact, and typically display more extreme values due to normalization. This process amplifies differences in the parametrization shapes, which are reflected by more significant differences between nSC, $\zeta$, $\Gamma$ and nASC for the bivariate and elliptic power distributions at higher values of $\rho$.

Experimentally, while there are constraints for the distribution of reference particle $v_n$ \cite{ATLAS:2013xzf}, there is currently little theoretical insight or experimental data to guide the development of parametrizations for $p(v'_n)$. For our toy model, we relied on the assumption that $v'_n$ and $v_n$ have the same parametrization, but different parameter values. There is no way to validate the quality of this assumption, and in reality, $p(v'_n)$ could be fundamentally different than what our model describes. However, when considering distributions that specifically are parametrized to reflect fluctuations in $v'_n$, the use of a Gaussian distribution for $p(v'_n)$ with a fixed $\sigma(v'_n)$ is a reasonable approximation, because even if the shape of the distribution $p(v'_n)$ is vastly different, it will still have a value for $\sigma(v'_n)$, and will display identical second order fluctuations to a Gaussian distribution parametrized by the same $\sigma(v'_n).$

The correlation model between $v'_n$ and $v_n$ dictates precisely how the joint fluctuations between $v'_n$ and $v_n$ are produced and evaluated by observables, and different choices for this correlation model will likely have a significant effect on the results of the analysis. The Gaussian copula \cite{Meyer_2013,Takeuchi_2010} was selected for this analysis because it allows for the direct specification of $\rho$, and simply reduces to the equation for a bivariate Gaussian distribution when each marginal is a Gaussian distribution \cite{Takeuchi_2010}. Generally, there is no reason that the correlations between $v'_n$ and $v_n$ should be described using a Gaussian copula model over any other type of correlation. However, regardless of the underlying distribution function of the correlation, the Pearson correlation coefficient $\rho(v'_n,v_n)$ representing second order fluctuations in $v_n$ and $v'_n$ will always exist for a given joint probability density $p(v'_n,v_n)$, as long as up to second order moments for each variable can be evaluated. While there are plenty of cases in which the correlation coefficient fails to accurately represent dependency between $v'_n$ and $v_n$, we can always produce a model that displays a given $\rho(v'_n,v_n)$ using the Gaussian copula, and the correlation coefficient is extremely common as a linear correlation measure for joint probability distributions.

\section{Discussion and Conclusion}\label{sec:6}

Multiparticle correlations are one of the most powerful and fundamental tools for studying the QGP produced in heavy ion collisions. In this study, for the first time, we describe how these multiparticle correlations can be harnessed to measure fluctuations in jet energy loss, among other phenomena. 

In this paper, we have developed a toy model to construct and evaluate distributions of $v_n$ and $v'_n$ with varying fluctuations in $v'_n$, and varying correlations between $v_n$ and $v'_n$. Using this toy model we characterize the sensitivity of four unique observables to fluctuations and correlations in $v'_n$ and $v_n$. While they depend on parametrizations and correlation models, we show how the sensitivities of each observable can be used to constrain values for fluctuations in hard probe azimuthal anisotropies, and how they relate to the fluctuations of the soft sector azimuthal anisotropies within heavy ion collisions. Greater accuracy can be obtained when constraining the correlations and fluctuations in $v'_n$ and $v_n$ using a simultaneous measurement of multiple observables. The method described in this paper presents a unique opportunity to constrain different theoretical models for jet energy loss processes and fluctuations by evaluating observables detailed in Sec.~\ref{observablesection} in both simulation and experimental settings. 
Theoretical calculations of the observables in this paper can create baselines for both the fluctuation and correlations between $v'_n$ and $v_n$. Futhermore, they can relate these fluctuations to hydrodynamical variables and jet energy loss models. Experimental measurements of these observables will provide concrete information about the fluctuations and correlations in $v'_n$ and $v_n$, as well as constraints on the associated physics processes characterized by theoretical work. 

Recent and ongoing high luminosity LHC runs, coupled with sPHENIX and STAR data from the upcoming AuAu RHIC run will provide unprecedented precision for jet measurements. This abundance of data will provide an opportunity to apply the observables and analysis method discussed in this paper, and isolate estimates for the fluctuations in jet energy loss within the QGP medium. 
Moreover, while we have focused primarily on jet POIs, our approach and observables can also be used with other types of POI to probe fluctuations in rare probe production and interactions with the QGP.

\section{Acknowledgements}

 A.H., A.M.S., and X. W. acknowledge support from National Science
Foundation Award Number 2111046. J.N.H. acknowledges the support from the US-DOE Nuclear Science Grant No.
DE-SC0023861 and within the framework of the Saturated Glue (SURGE) Topical Theory
Collaboration

\clearpage

\bibliographystyle{unsrturl}
\bibliography{refs,refs_noinspire}

\appendix
\section{Evaluation of Multiparticle correlations}\label{ap:Q_vectors}

 The $Q$-Cumulant method \cite{Bilandzic:2010jr,Bilandzic:2012wva} is a more computationally efficient method used to measure angular particle correlations without double counting particles, and ignoring contributions from higher harmonics. The $Q$-Cumulant method writes the even moments of $v_n$, $\langle v_n^{2k}\rangle$, as a function of event $Q_n$ vectors with weights multiplied by different powers $r$, a sum over all $N$ reference particles in an event:
\begin{equation}
 Q_{n,r} = \sum_{i} w_i^r e^{in\phi_i}
\end{equation}
 where $\phi_i$ indicates the azimuthal angle of each particle.
 We include a weighting $w_i$ for each particle, which is raised to the $r$ power. To incorporate the azimuthal anisotropies displayed by jets, or any other differential particle species, we use the "reduced" vector: 
 \begin{equation}\label{littleq}
 q_{n,r} = \sum_{j} w_j^r e^{in\psi_j}.
\end{equation}
 where $j$ is an index for for each differential particle, featuring azimuthal angle $\psi.$ 

For this analysis we also assume there is no overlap between the POI and the reference particles, as specifically jets and soft particles will have significantly different selection criteria.

 \subsection{Two Particle Correlations}

Following the formalism from \cite{Bilandzic:2013kga,Holtermann:2023vwr}, we obtain expressions for the two-particle and four-particle correlations used to construct the observables discussed in this paper. The expressions are arrived at by evaluating the overlap between particles, as illustrated by a derivation of the expression for $\avg{{v'}_n^2}$, 
\begin{eqnarray}
  \avg{{v'}_n^2} &= \avg{e^{in(\psi_1 - \psi_2)}}
  &= \frac{\sum_{k = 1}^{N}\sum_{j=1, j\neq i}^{N} w_k w_j e^{in(\phi_k - \phi_j)}}{\sum_{k = 1}^{N}\sum_{j=1, j\neq i}^{N} w_k w_j}.
\end{eqnarray}
Then, we note that the iterated sum over unique pairs of particles can be written as a subtraction of all duplicates ($\psi_a$,$\psi_a$) from every possible combination of angle pairs (note that double counting of angle pairs $(\psi_a,\psi_b), (\psi_b,\psi_a)$ is accounted for in the denominator of the fraction): 

 \begin{eqnarray}
   \frac{\sum_{k = 1}^{N}\sum_{j=1, j\neq i}^{N} w_k w_j e^{in(\psi_k - \psi_j)}}{\sum_{k = 1}^{N}\sum_{j=1, j\neq i}^{N} w_k w_j} = \frac{\sum_{k=1}^{N}\sum_{j=1}^{N} w_jw_k e^{in(\psi_k-\psi_j)} - \sum_{l=1}^{N} w_l^2 e^{in(\psi_l - \psi_l)}}{\sum_{k=1}^{N}\sum_{j=1}^{N} w_jw_k - \sum_{l=1}^{N} w_l^2}.
\end{eqnarray}

 Using the definition of $q_{n,r}$ given in Eq.~(\ref{littleq}), we obtain: 

 \begin{eqnarray}
   \frac{\sum_{k=1}^{N}\sum_{j=1}^{N} w_jw_k e^{in(\psi_k-\psi_j)} - \sum_{l=1}^{N} w_l^2 e^{in(\psi_l - \psi_l)}}{\sum_{k=1}^{N}\sum_{j=1}^{N} w_jw_k - \sum_{l=1}^{N} w_l^2} = \frac{q_{n,1}q_{n,-1} - q_{0,2}}{q_{0,1}^2 - q_{0,2}}.
\end{eqnarray}
Following this procedure for each two-particle and four-particle correlation in Table \ref{correlators}, we present their form as a function of the $q_{n,r}$ and $Q_{n,r}$ vectors. Note that all averages here, indicated by angle brackets, indicate an average taken over an ensemble of events.

\begin{equation}
 \avg{v_n^2} = 
 \avg{\frac{|Q_{n,1}|^2 - Q_{0,2}}{|Q_{0,1}|^2 - Q_{0,2}}}
\end{equation}

\begin{equation}
 \avg{V'_nV_n^*} = \avg{\frac{q_{n,1}Q_{n,1}^*}{q_{0,1}Q_{0,1}^*}}
\end{equation}

\begin{equation}
 \avg{{v'}_n^2} = \avg{\frac{|q_{n,1}|^2 - q_{0,2}}{|q_{0,1}|^2 - q_{0,2}}}
\end{equation}

\subsection{Four-Particle Correlations}

Following a similar process as for the two particle correlations, and described in detail in Ref.~\cite{Holtermann:2023vwr}, we obtain the following expression for the various four-particle correlations using $Q_n$ vectors: 

\begin{equation}
\begin{split}
 \avg{v_n^4} &= \left\langle \left(\left|Q_{n, 1}\right|^4-Q_{2 n, 2} Q_{n, 1}^{* 2}-4 Q_{0,2}\left|Q_{n, 1}\right|^2-Q_{2 n, 2}^* Q_{n, 1}^2+|Q_{2 n, 2}|^2+2\left|Q_{0,2}\right|^2+ \right.\right.\\
 &\left.\left.+4 Q_{n, 3} Q_{n,1}^* + 4 Q_{n, 3}^* Q_{n,1} - 6Q_{0,4} \right)\right.\\
 &\times \left.\left(Q_{0,1}^4-6 Q_{0,1}^2 Q_{0,2}+3 Q_{0,2}^2+8 Q_{0,1} Q_{0,3}-6 Q_{0,4}\right)^{-1}\right\rangle \\
\end{split}
\end{equation}

\begin{equation}
\begin{split}
 \avg{V'_nV_n^* v_n^2} &= \left\langle \left( q_{n,1}Q_{n,1}^*|Q_{n,1}|^2 - 2q_{n,1}Q_{n,1}^*Q_{0,2} - q_{n,1}Q_{n,1}Q_{2n,2}^*+2q_{n,1}Q_{n,3}^* \right)\right. \\
 &\times \left.\left(q_{0,1}Q_{0,1}^3 - 3q_{0,1}Q_{0,1}Q_{0,2} + 2q_{0,1}Q_{0,3} \right)^{-1}\right\rangle
 \end{split}
\end{equation}

\begin{equation}
\begin{split}
 \avg{(v^\prime)_n^2 v_n^2} &= \left\langle \left( |q_{n,1}|^2|Q_{n,1}|^2 - q_{0,2}|Q_{n,1}|^2 -|q_{n,1}|^2 Q_{0,2} + q_{0,2}Q_{0,2} \right)\right. \\
 &\times \left.\left(q_{0,1}^2Q_{0,1}^2 - q_{0,2}Q_{0,1}^2 - q_{0,1}^2Q_{0,2}+q_{0,2}Q_{0,2}\right)^{-1}\right\rangle
 \end{split}
\end{equation}

\begin{equation}
\begin{split}
 \avg{(V'_nV_n^*)^2} &= \left\langle \left( 
 q_{n,1}^2Q_{n,1}^{* 2} - q_{2n,2}Q_{n,1}^{* 2} - q_{n,1}^2Q_{2n,2}^{*} + q_{2n,2}Q_{2n,2}^*\right)\right. \\
 &\times \left.\left(q_{0,1}^2Q_{0,1}^{* 2} - q_{0,2}Q_{0,1}^{* 2} - q_{0,1}^2Q_{0,2}^{*} + q_{0,2}Q_{0,2}^*\right)^{-1}\right\rangle.
 \end{split}
\end{equation}

\section{ Analytic formulas for joint moments of a bivariate Gaussian Distribution }\label{ap:Calc}

 The probability density function for two arbitrary stochastic variables $X$ and $Y$ parametrized with a bivariate Gaussian distribution with parameters ($\mu_x$, $\mu_y$, $\sigma_x$, $\sigma_y$, $\rho$) is given by: 
 \begin{equation}
  p(X,Y) = \frac{1}{2\pi \sigma_x \sigma_y \sqrt{1-\rho^2}}\exp\left(\frac{1}{2(1-\rho)^2}\left(\frac{(X-\mu_x)^2}{\sigma_x^2} + \frac{(Y-\mu_y)^2}{\sigma_y^2} - \frac{2\rho(X-\mu_x)(Y-\mu_y)}{\sigma(v_n)\sigma(v'_n)}\right)\right)
 \end{equation}
The moment generating function is used to evaluate the moments of a distribution. Generally for a two variable probability distribution, the moment generating function is a function of two dummy variables, $t_x$ and $t_y$, and is written as follows 
\begin{equation}
 M_{X,Y}(t_x,t_y) = \avg{e^{t_x X+ t_y Y} } = \iint e^{t_x X+ t_y Y} p(X,Y) dX dY
\end{equation}
where in this case, an average indicated by angle brackets is an integral over all values of $X$ and $Y$. For a bivariate Gaussian distribution the moment generating function has the form: 
\begin{equation}
M_{X,Y}(t_x,t_y) = \exp{ \left(\mu_x t_x + \mu_y t_y + \frac{1}{2}\sigma_x^2 t_x^2 + \frac{1}{2}\sigma_y^2 t_y^2 + \rho \sigma_x \sigma_y t_x t_y\right)}
\end{equation}
Finally, the moments are extracted from the moment generating function by evaluating terms in the Taylor expansion of $M_{X,Y}(t_x,t_y)$ around the origin as follows:

\begin{equation}
 \avg{X^nY^m} = \frac{\partial^{n+m} M_{XY}(t_x,t_y)}{\partial X^n \partial Y^n}\Bigg|_{t_x = t_y = 0}.
\end{equation}

For a concise example, we evaluate $\avg{XY}$ using this method.

\begin{equation}
 \avg{XY} = \frac{\partial^2 M_{XY}(t_x,t_y)}{\partial X\partial Y}\Bigg|_{t_x = t_y = 0} = \frac{\partial}{\partial Y} \frac{\partial M_{XY}(t_x,t_y)}{\partial x}\Bigg|_{t_x = t_y = 0}
  = \frac{\partial}{\partial Y}\left( M_{XY}(t_x,t_y)\left[\mu_x + \sigma_x^2 t_x + \rho \sigma_x\sigma_y t_y \right] \right)\Bigg|_{t_x = t_y = 0}
\end{equation}
\begin{equation}
 = M_{XY}(t_x,t_y)\left[\mu_x + \sigma_x^2 t_x + \rho \sigma_x\sigma_y t_y\right]\left[\mu_y + \sigma_y^2 t_y + \rho \sigma_x \sigma_y t_x\right]+ M_{XY}(t_x,t_y)\left[\rho \sigma_x\sigma_y\right]\Bigg|_{t_x = t_y = 0}
\end{equation}
Then, we evaluate at $t_x = t_y = 0$: 
\begin{equation}
 \avg{XY}= \frac{\partial^2 M_{X,Y}(0,0)}{\partial x \partial y} = \mu_x\mu_y + \rho \sigma_x\sigma_y
\end{equation}

Using this procedure, we derive formulas for the moments used in this paper up to four particles (detailed in Table~\Ref{correlators}). 

\begin{eqnarray}
 \avg{X^2} &=& \mu_x^2 + \sigma_x^2 \\
 \avg{Y^2} &=& \mu_y^2 + \sigma_y^2 \\
 \avg{XY} &=& \mu_x\mu_y + \rho \sigma_x \sigma_y\\
 \avg{X^4} &=& \mu_x^4 + 6\sigma_x^2 \mu_x^2 + 3 \sigma_x^4\\
 \avg{X^3Y} &=& \mu_x^3\mu_y + 3\mu_x^2 \rho \sigma_x\sigma_y + 3 \sigma_x^2 \mu_x \mu_y + 3 \sigma_x^2 \rho \sigma_x \sigma_y\\
 \avg{X^2Y^2} &=& \mu_x^2 \mu_y^2 + 4 \rho \sigma_x\sigma_y \mu_x \mu_y + \mu_x^2 \sigma_y^2 + \sigma_y^2 \mu_x^2 + \sigma_x^2\sigma_y^2 + 2\rho^2 \sigma_x^2\sigma_y^2
\end{eqnarray}

\end{document}